\documentclass[11pt]{article}
\usepackage{fullpage}
\usepackage{graphicx}
\usepackage{amsmath}

\usepackage{amssymb}
\usepackage[top=1in,bottom=1in,left=1in,right=1in]{geometry}

\usepackage{amsbsy}
\usepackage{mathtools}

\usepackage{natbib}
\usepackage{url}

\usepackage{multirow}
\usepackage{amsbsy}
\usepackage{mathtools}

\usepackage{microtype}

\usepackage[table]{xcolor}
\usepackage{hhline}

\providecommand{\abs}[1]{\lvert#1\rvert}

\makeatletter
  \newcommand\scripty{\@setfontsize\scripty{6pt}{8}}
\makeatother

\makeatletter
  \newcommand\script7{\@setfontsize\scripty{7pt}{8}}
\makeatother

\usepackage{url}

\begin{document}


\title{Multiple network alignment via multiMAGNA++}

\author{Vipin Vijayan and Tijana Milenkovi\'{c}\footnote{To whom correspondence should be addressed}\\Department of Computer Science and Engineering, ECK Institute for Global Health,\\and Interdisciplinary Center for Network Science and Applications (iCeNSA)\\University of Notre Dame, Notre Dame, IN 46556, USA}

%
\date{}

\maketitle

\begin{abstract}
\noindent
{\bf Motivation: }Network alignment (NA) aims to find a node
mapping between molecular networks of different species that
identifies topologically or functionally similar network
regions. Analogous to genomic sequence alignment, NA can be used to
transfer biological knowledge from well- to poorly-studied species
between aligned network regions. Pairwise NA (PNA) finds similar
regions between two networks while multiple NA (MNA) can align more
than two networks. We focus on MNA. Existing MNA methods aim to
maximize total similarity over all aligned nodes (node
conservation). Then, they evaluate alignment quality by measuring the
amount of conserved edges, but only after the alignment is
constructed. Directly optimizing edge conservation during alignment
construction in addition to node conservation may result
in superior alignments. 
\\
{\bf Results: }Thus, we present a novel MNA approach called multiMAGNA++ 
that can achieve this. Indeed, multiMAGNA++ generally outperforms or
is on par with the existing MNA methods, while often completing faster
than the existing methods. That is, multiMAGNA++ scales well to larger
network data and can be parallelized effectively. During method
evaluation, we also introduce new MNA quality measures to allow for
more complete alignment characterization as well as more fair MNA
method comparison compared to using only the existing alignment
quality measures.\\
{\bf Code availability: }Code is available upon request.\\ 
{\bf Contact: }tmilenko@nd.edu
\end{abstract}


\section{Introduction}
\label{sec:intro}

Networks can model a variety of real-world systems, including
biological ones.  Many networks are relevant to biological systems,
such as protein-protein interaction (PPI)
\citep{BioGRID}, gene co-expression  \citep{GCO0}, or metabolic \citep{METABOL0} networks.  In  PPI networks, nodes are proteins and
edges are physical interactions between the proteins.  The
interactions between the proteins control and carry out complex
cellular functions, and hence, studying PPI networks is important.
Biotechnological advances have made PPI network data available for
many species \citep{BioGRID}.  Functions of many proteins in many
species remain unknown \citep{NAtransfer1,NAtransfer2}, and hence,
there is need for across-species transfer of existing functional
knowledge from well-studied species to poorly-studied ones.
Typically, genomic sequence alignment has been used for this purpose
\citep{BLAST}.  However, since proteins interact in the cell to
carry out cellular function and since PPI networks model these
interactions (whereas sequence alignment studies genes in isolation),
biological network alignment (NA) can be used for this knowledge
transfer too, in order to complement the biological insights that have
already been gained via sequence alignment.

NA aims to find a node mapping between networks that identifies
topologically or functionally similar network regions.  As such, NA
can be used to transfer biological knowledge across species between
the species' conserved (aligned) PPI network regions
\citep{NAtransfer0}.  While we focus on NA of PPI networks and on the
domain of computational biology
\citep{SharanReview,ClarkReview,ConeReview}, NA is applicable to any
network type (e.g., gene regulatory or metabolic networks) and a wide
number of fields (e.g., pattern recognition
\citep{NApaperConte,NApaperForsyth}, language processing
\citep{NApaperBayati}, social networks
\citep{NApaperKoutra,NApaperKollias}, and computer vision
\citep{NApaperDuchenne}).

NA is related to the subgraph isomorphism problem, which asks whether
one network (or graph) is an exact subgraph of another network.  The
subgraph isomorphism problem is NP-complete, meaning that there is no
efficient method to solve it exactly for large input networks.  NA is
a more general problem as it asks how to best ``fit'' one network into
another network, even if the first network is not an exact subgraph of
the second one.  It is not obvious how to measure the quality of the
fit of one network into another one.  But a measure that is widely
used quantifies the amount of conserved (aligned) edges, or in other
words, the size of the common conserved subgraph between the aligned
networks.  Maximizing edge conservation is  NP-hard 
\citep{MIGRAAL}.  Thus, heuristic methods need to be sought for NA.

NA can be either {\em local} or {\em global} (see
\cite{LocalVsGlobal,ConeReview,ElmsallatiReview} for a review).  
Initial NA focus was on local NA, which finds smaller, highly
conserved regions, e.g., biological pathways or protein complexes,
among networks.  However, local NA does not generally capture large
conserved subgraphs, and the aligned regions can overlap, leading to
an ambiguous (from a mathematical perspective) many-to-many node
mapping.  More recent efforts have focused on global NA, which maps
entire networks to each other.  Since global NA can find large
conserved subgraphs shared by the aligned networks, and since large
conserved subgraphs have been argued to be more helpful in aiding
transfer of knowledge between networks
\citep{HGRAAL,GHOST}, we focus on global NA.

NA can be classified in another way: as pairwise NA (PNA) or multiple
NA (MNA).  PNA finds similar regions between two networks while MNA
can align more than two networks.  PNA typically produces an injective
(one-to-one) node mapping between two networks, which results in
aligned node {\em pairs}.  MNA produces an alignment consisting of
aligned node {\em clusters}.  Given an alignment of multiple networks,
if an aligned cluster contains more than one node from a single
network, it is a many-to-many MNA.  If there is a maximum of one node
per network in every aligned cluster, it is a one-to-one MNA.  While
MNA may lead to deeper biological insights compared to PNA since it
captures at once functional knowledge that is common to multiple
species, which is why we focus on MNA, MNA is computationally much
harder than PNA since the complexity of the NA problem increases
exponentially with the number of networks to be aligned.

Existing global NA methods are typically {\em two-stage} aligners:
they first calculate similarity (with respect to some node cost
function) between nodes from different networks, and then they use an
alignment strategy to identify high scoring alignments with respect to
the total similarity over all aligned nodes (also known as node
conservation).  Examples of two-stage {\em PNA} methods are IsoRank
\citep{IsoRank}, GHOST \citep{GHOST}, and the GRAAL family of methods
\citep{GRAAL,HGRAAL,MIGRAAL}.  Examples of two-stage {\em MNA}
methods are IsoRankN \citep{IsoRankN}, MI-Iso
\citep{Faisal2014}, SMETANA \citep{SMETANA}, BEAMS \citep{BEAMS}, NetCoffee \citep{NetCoffee}, CSRW \citep{RandomWalk}, and FUSE \citep{FUSE}. For an overview of these methods, see Section
\ref{sec:supplintro}.

\citet{Faisal2014} and \citet{FairEval} pointed out a key issue with 
two-stage NA methods. Namely, they evaluated these methods by mixing
and matching their node cost functions and alignment strategies.  They
showed that node cost function of one method and alignment strategy of
another method can (and typically do) yield a new superior method.
This finding highlighted the need for properly evaluating a new
two-stage NA method against the existing ones, by using the above
mix-and-match strategy, in order to determine whether it is the new
method's node cost function or its alignment strategy (or both) that
leads to its potential superiority.

Another important issue that exists with the two-stage NA methods is
as follows.  Once the two-stage NA methods generate an alignment that
has high node conservation, they typically evaluate the quality of the
alignment using some other measure that is different from the node
conservation measure used to guide the alignment construction.  As
already noted, they evaluate alignment quality with respect to the
amount of conserved edges.  That is, the two-stage methods align
similar nodes between networks hoping to conserve many edges, but they
calculate the amount of conserved edges only {\it after} the alignment
is constructed. Even the two-stage NA methods that optimize the best measures of node conservation, which are based on topological similarity of {\em extended} network neighborhoods of nodes in question \citep{Faisal2014,FairEval}, cannot increase edge conservation directly.

To address this issue, we recently introduced MAGNA \citep{MAGNA} to
directly optimize edge conservation {\it while} the alignment is
constructed.  MAGNA is a {\em search-based} (rather than a two-stage)
PNA approach.  Search-based aligners can directly optimize edge
conservation or any other alignment quality measure.  Intuitively,
MAGNA uses a novel crossover function, which creates a child alignment
by combining two parent alignments, and a genetic algorithm, in order
to simulate a population of alignments that evolve over multiple
generations.  We note that we used a genetic algorithm within MAGNA
simply as a proof of concept that even such a simple heuristic
algorithm, when used to directly optimize edge conservation during
alignment construction, would result in superior alignments when
compared to two-stage network aligners. Using a more advanced approach
instead of a genetic algorithm would likely even further improve
alignment quality.  Indeed, in systematic evaluations against
state-of-the-art two-stage methods (IsoRank, MI-GRAAL, and GHOST), on
networks with known and unknown node mappings,
MAGNA outperformed
all of the existing methods, in terms of both node and edge
conservation as well as both topological and functional alignment
accuracy.  Importantly, in addition to constructing its own superior
alignments from scratch, owing to its powerful crossover function,
MAGNA can combine alignments of existing methods to further improve
them.  Because simultaneously maximizing both node and edge
conservation could further improve alignment quality
\citep{GREAT,WAVE,NETAL}, even more recently we extended MAGNA into a
new MAGNA++ PNA framework \citep{MAGNA++}.  Indeed, when we used
MAGNA++ to optimize both node and edge conservation, we improved
alignment quality compared to optimizing node conservation only (as
existing two-stage aligners do) or edge conservation only (as MAGNA
does).  Additional search-based PNA methods that have appeared in
parallel to or since MAGNA++ are NABEECO \citep{NABEECO}, GEDEVO
\citep{GEDEVO}, and Optnetalign \citep{Optnetalign}.

In this paper, we introduce multiMAGNA++, an extension of MAGNA++ from
PNA to MNA.  That is, we propose multiMAGNA++ as a novel global
one-to-one MNA algorithm.  We want to show that directly optimizing
edge conservation in addition to node conservation using
a genetic algorithm as a proof-of-concept results in superior
alignments for MNA compared to MNA algorithms that optimize node
conservation only.  Like MAGNA++, multiMAGNA++ is a search-based
method that directly optimizes both edge and node conservation while
the alignment is constructed.  The key computational novelties of
multiMAGNA++ compared to MAGNA++ are (i) our representation of an MNA
using permutations, and (ii) a new crossover function for producing
child alignments from parent alignments that allows for aligning
multiple networks (unlike the crossover function of MAGNA++, which
allows for aligning only two networks).

Of the existing MNA methods, all are two-stage aligners except
GEDEVO-M, where the latter is a search-based MNA equivalent of
PNA-based GEDEVO.  In evaluations against the existing MNA methods
(IsoRankN, MI-Iso, GEDEVO-M, BEAMS, and FUSE), multiMAGNA++ overall
outperforms or is comparable to the existing methods with respect to
multiple alignment quality measures and on multiple datasets.  In the
process of method evaluation, we also introduce new alignment quality
measures for MNA, to allow for more fair method comparisons compared
to using only the existing alignment quality measures.

The paper is organized as follows.  In Section
\ref{sec:magna}, we describe our new multiMAGNA++.  In
Section \ref{sec:eval}, we describe how we evaluate multiMAGNA++
against the existing MNA methods.  In Section \ref{sec:results}, we
discuss our results.  In Section
\ref{sec:conclusion}, we conclude our study.


\section{Methods}
\label{sec:methods}

\subsection{MultiMAGNA++}
\label{sec:magna}

MultiMAGNA++ is a genetic algorithm (GA) \citep{Genetic} that
maximizes an alignment quality measure by evolving a population of
alignments over time.  We use a GA only as a proof-of-concept that
directly optimizing edge conservation in addition to node
conservation would result in superior alignments when compared to the
existing methods that optimize node conservation only.  Below, we
introduce GA-related terminology needed to understand multiMAGNA++
(Section \ref{sec:geneticalgo}), our alignment representation of an
MNA using permutations (Section \ref{sec:representation}), our novel
crossover function that relies on the permutation-based MNA
representation for producing a new child alignment from parent
alignments (Section \ref{sec:cross}), and our fitness function that we
optimize while crossing and evolving alignments (Section
\ref{sec:fitness}). Then, we describe multiMAGNA++'s parameter values 
 (Section \ref{sec:params}) and its time complexity (Section
 \ref{sec:complexity}).

\subsubsection{Genetic algorithm (GA)}
\label{sec:geneticalgo}

A GA is a search heuristic that optimizes a {\em fitness function} (in
our case, an alignment quality measure) using a population of {\em
members} (in our case, alignments).  Associated with each member is
its {\em fitness value}, which is calculated using the fitness
function.  Beginning with a population of members for the first
generation, the GA creates a new population in every generation by
keeping an {\em elite fraction} of the population (i.e., the fraction
of the population with the best fitness) from the previous generation
and filling the remainder of the population with members produced by
{\em crossovers}.  A crossover of two {\em parent} members produces a
new {\em child} member that ideally resembles both of the parents.
The GA selects parent members for crossover using a {\em selection
algorithm} that chooses parents from the population of members with
probability in proportion to the members' fitness.  Since the GA keeps
the elite members of each generation, the optimal fitness does not
decrease from one generation to the next.  As the GA produces newer
generations, the optimal fitness will ideally increase until it
reaches a {\em stopping criterion}.  We take the fittest member from
the final generation as the result of the optimization process.

\subsubsection{Our representation of an MNA}
\label{sec:representation}

To calculate the fitness of an alignment and the crossover of two
alignments, the GA requires an explicit representation of an
alignment.  MAGNA++ represents a PNA of two networks using a single
permutation.  For multiMAGNA++, we extend this to represent an MNA of
$k$ networks using $k-1$ permutations.

First, we describe how to represent a PNA using a single permutation.
Let $G_1(V_1,E_1)$ and $G_2(V_2,E_2)$ be two networks, with node and
edge sets $V_l$ and $E_l$, respectively, where $l=1,2$.  Without loss
of generality, assume that $m \leq n$, where $m = \abs{V_1}$ and $n =
\abs{V_2}$.  A PNA of $G_1$ to $G_2$ is a total injective mapping $f
\colon V_1 \mapsto V_2$; that is, every element in $V_1$ is matched
uniquely with an element in $V_2$.  If $m = n$, then $f$ is a
bijective mapping.  We need this constraint of $m = n$ to be satisfied
in order to be able to represent a PNA as a permutation (as described
below), and we can easily achieve this without making any special
assumptions (Section \ref{sec:supplrepresentation}).
Given the above definitions, the resulting mapping $f$ is a set of
aligned pairs $\{(v,f(v)) \,|\, v \in V_1\}$.  Now, how to represent
$f$ using a permutation?  A permutation is a bijective mapping between
two sets of integers: $\{1,2,\ldots,n\}$ and $\{1,2,\ldots,n\}$.
Given this, and given that $f$ is a bijective mapping between nodes of
two networks, $f$ can be represented as a permutation by fixing the
ordering of the nodes in each of the two networks \citep{MAGNA}.  For
this reason, henceforth, we refer to a PNA, an injective mapping, and
a permutation as synonyms.

\begin{figure}
\centering
\includegraphics[width=0.54\linewidth]{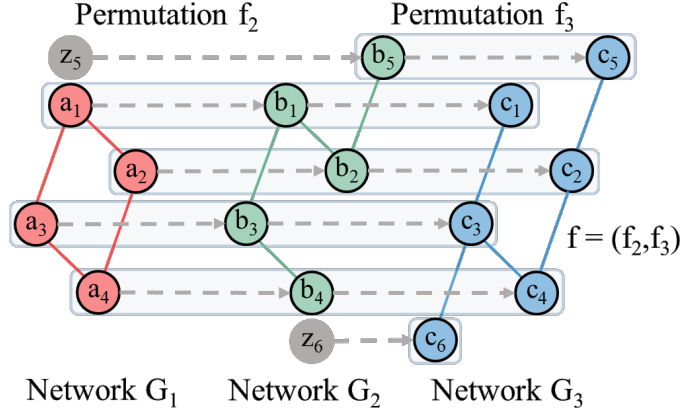}
\caption{An example of how an MNA of three networks is represented using two permutations. Networks $G_1(V_1,E_1),G_2(V_2,E_2)$, and $G_3(V_3,E_3)$ are ordered in terms of the number of nodes from the smallest to the largest one. Since $\abs{V_2} = 5$ and $\abs{V_3} = 6$, $f_2$ and $f_3$ are permutations of size 5 and 6, respectively. (Here, nodes $z_5$ and $z_6$ are dummy nodes added to $V_1$ and $V_2$, respectively, intended to enforce the $m = n$ constraint described in the text, i.e., to ensure that $\abs{{\bar V_1}} = \abs{V_2} = 5$ and $\abs{{\bar V_2}} = \abs{V_3} = 6$; Section S2.1.1). The permutations create disjoint clusters in this toy example that cover all the nodes in the networks. For example, the cluster created through the mapping $a_1 \mapsto b_1 \mapsto c_1$ is $(\{a_1\},\{b_1\},\{c_1\})$, and the cluster created through the mapping $b_5 \mapsto c_5$ is $(\varnothing, \{b_5\}, \{c_5\})$.
}
\label{fig:compgraph}
\end{figure}

Second, we describe how we extend the above notion to allow for
representing an MNA of $k$ networks using $k-1$ permutations.  Let
$G_1(V_1,E_1)$, $G_2(V_2,E_2)$, $\ldots$, $G_k(V_k,E_k)$ be $k$
networks, with node and edge sets $V_l$ and $E_l$, respectively, where
$l=1,\ldots,k$.  Without loss of generality, assume that the networks
are ordered in terms of the number of nodes from the smallest to the
largest one.  A one-to-one MNA of $k$ networks is a set of disjoint
clusters where each cluster is represented as a tuple
$(a_1,a_2,\ldots,a_k)$, such that: (i) $a_l \subseteq V_l$, (ii) $a_l
\cap b_l = \varnothing$ for two different clusters
$(a_1,a_2,\ldots,a_k)$ and $(b_1,b_2,\ldots,b_k)$, and (iii)
$\abs{a_l} \leq 1$, for $l = 1,\ldots,k$.  That is, a one-to-one MNA
is a set of disjoint clusters, each of which can contain at most one
node from each network.  On the other hand, if we omit the third
condition above, so that $\abs{a_l}$ can be larger than 1, then the
clusters would form a many-to-many (rather than one-to-one) MNA.
However, the focus of our work is on one-to-one MNA, and henceforth,
we refer to such an alignment simply as an MNA.  Now, how to represent
an MNA of $k$ networks using $k-1$ permutations?  We achieve this as
follows.  We represent an MNA using permutations $f_2,\ldots,f_{k}$,
which are bijective mappings between pairs of networks that are
adjacent when ordered by size, such that $V_{1} \xmapsto{f_2} V_{2}
\xmapsto{f_3} \ldots \xmapsto{f_{k-1}} V_{k-1} \xmapsto{f_k} V_k$.
The permutations correspond to a set of disjoint node clusters that
cover (not necessarily all) nodes in the $k$ networks (Figure
\ref{fig:compgraph}).  The cluster set can be denoted as
$\{(a_1,a_2,\ldots,a_k) \,|\, a_l = a_l(v), \, l=1,\ldots,k, \, v \in
V_k \}$, where $a_l(v)$ is defined as (i) $a_l(v) = \{v\} \mbox{ if }
l = k$, (ii) $a_l(v) = \{u\} \mbox{ if } l < k \mbox{ and } f_{l+1}(u)
\in a_{l+1} \mbox{ for some } u \in V_l$, and (iii) $a_l(v) =
\varnothing \mbox{ otherwise}$.  So, an MNA of $k$ networks can be
represented using a tuple $f$ of $k-1$ permutations, $f =
(f_2,\ldots,f_k)$, which we call a multi-permutation. Thus,
henceforth, we refer to an MNA and a multi-permutation as synonyms.

\subsubsection{Crossover function}
\label{sec:cross}

The core of a GA is the crossover function, which creates a child
alignment from two parent alignments \citep{Genetic}.  A crossover
function such that the child alignment has characteristics of both
parents results in better GA performance.  MAGNA++ made use of a novel
crossover function for PNA, which used the concept of a Cayley graph
to create a child permutation (see below).  Here, we extend MAGNA++'s
crossover function by formulating a notion of a Cayley graph for MNA.

First, we describe MAGNA++'s crossover function using its notion of a
Cayley graph in the context of PNA.  Recall that we can represent a
PNA of $G_1$ to $G_2$ using a permutation $f$ of size $n$.  Let $S_n$
denote the set of all permutations of size $n$.  A transposition of a
permutation is a new permutation that fixes every element of the
original permutation except two elements, which are swapped.  The
transposition of a permutation and the original permutation are of
similar alignment quality since a small perturbation of a permutation
is not expected to greatly affect alignment quality.  Thus, in order
to design a crossover function, a graph can be created whose topology
takes advantage of the fact that a transposition of a permutation does
not greatly affect its alignment quality.  MAGNA++ constructs such a
graph so that its node set is $S_n$ and so that there is an edge
between nodes $f \in S_n$ and $g \in S_n$ if and only if there is a
transposition $\nu$ such that $f = \nu g$, i.e. if and only if the
permutations $f$ and $g$ differ by a transposition.  The resulting
graph is the PNA-based Cayley graph.  This graph has the desired
property that, given permutations $f$ and $g$, the child permutation
$f \otimes g$ will be the midpoint of the geodesic (shortest) path
between $f$ and $g$, and as such, it is expected to share
approximately half of its aligned pairs with each of $f$ and $g$
\citep{MAGNA}.

Second, we define our crossover function using the formulation of the
Cayley graph in the context of MNA.  Let $n_l = \abs{V_l}$ for $l =
1,\ldots,k$ and let $S_{n_l}$ denote the set of all permutations of
size $n_l$.  Recall that we can represent an MNA of $k$ networks using
a multi-permutation $f = (f_2,f_3,\ldots,f_k)$ containing permutations
of sizes $n_2,n_3,\ldots,n_k,$ respectively.  Thus, $S = S_{n_2}
\times S_{n_3} \times \ldots \times S_{n_{k}}$ is the set of all
multi-permutations (i.e., MNAs) of the $k$ networks.  Analogous to the
PNA-based Cayley graph above, we can create a graph whose topology
takes advantage of the fact that a transposition of a single
permutation in a multi-permutation does not greatly affect its
alignment quality.  We construct such a graph so that its node set is
$S$ and so that there is an edge between $(f_2,\ldots,f_{k}) \in S$
and $(g_2,\ldots,g_{k}) \in S$ if and only if there is a transposition
$\nu$ such that $f_{l'} = \nu g_{l'}$ for some $l' \in \{2,\ldots,k\}$
and $f_l = g_l$ for $l \ne l'$, i.e., if and only if only one pair of
permutations in the two multi-permutations differs only by a
transposition.  The resulting graph is the MNA-based Cayley graph.
This graph allows us to design a crossover function as follows.  Given
multi-permutations $f \in S$ and $g \in S$, let $f \otimes g$ be the
midpoint of the geodesic path between $f$ and $g$.  The permutations
in a multi-permutation are independent of each other in the sense that
changing one permutation does not affect another permutation.  So, $f
\otimes g = (f_2 \otimes g_2,f_3 \otimes g_3,\ldots,f_{k} \otimes
g_{k})$, where $f_l \otimes g_l$ is MAGNA++'s crossover function for
two permutations.  $f \otimes g$ will share characteristics with both
$f$ and $g$ since each permutation $f_l \otimes g_l$ in the
multi-permutation $f \otimes g$ will have characteristics of both
permutations $f_l$ and $g_l$ for $l=1,\ldots,k$.  Thus, if we let the
midpoint $f \otimes g$ be the crossover, then we can expect that the
child MNA shares characteristics of each of the two parent MNAs.

\subsubsection{Fitness function}
\label{sec:fitness}

The fitness function that multiMAGNA++ directly optimizes is a combined measure of both edge and node conservation that we define below. We
optimize this combined measure intentionally, because we
aim to find out whether directly optimizing edge conservation in addition to node conservation would result in superior
alignments when compared to the existing methods, most of which
optimize node conservation only (but then they evaluate the resulting
alignments based on their edge conservation; Section \ref{sec:intro}).
The alignment quality measure that we use is a convex combination of
an edge conservation measure, $S_E$, and a node conservation measure,
$S_N$: $\alpha S_E + (1-\alpha) S_N$.  The $\alpha$ parameter varies
from $0$ to $1$ and controls for the influence of edge vs. node
conservation. 
The edge and node conservation measures that we use are as follows.

The edge conservation measure that we use is conserved interaction
quality (CIQ) \citep{BEAMS}.  CIQ is a weighted sum of edge conservation
between all pairs of aligned clusters.  It is a generalization of the
established S$^3$
\citep{MAGNA} edge conservation measure from PNA to MNA.  CIQ is
calculated as follows.  Given clusters $a$ and $b$, let
$\abs{E_{a,b}}$ be the number of edges that connect the clusters.  Let
$r(a,b)$ be the number of networks that the edges which connect the
clusters belong to.  Let $s(a,b)$ be the number of networks that
contain at least one node in both clusters.  Let the edge conservation
between two clusters be $cs(a,b)$, where either (i) $cs(a,b) = 0
\mbox{ if } r(a,b) \le 1$ or (ii) $cs(a,b) =
\frac{r(a,b)}{s(a,b)} \mbox{ otherwise}$.  That is, $cs(a,b)$ is 0 if
no edges connect the two clusters or if the edges that connect the
clusters belong to only one network.  Otherwise, $cs(a,b)$ is the
fraction of networks that the edges connecting the two clusters belong
to.  Given edge conservation between all pairs of clusters, total edge
conservation is $S_E = \mbox{CIQ} = \frac{ \sum_{a,b} \abs{E_{a,b}}
cs(a,b) } { \sum_{a,b} \abs{E_{a,b}} }$.

The node conservation measure that we use is an MNA quality measure
that we propose as our contribution.  Node conservation refers to
internal cluster quality, meaning that in a good alignment, nodes in
each cluster should be highly similar to each other with respect to
some node cost function -- see below).  We measure a cluster's
internal quality as mean node similarity across all node pairs in the
cluster.  To account for all clusters, we take the mean of the above
measure across all clusters.  Formally, let $s(u,v)$ be the similarity
between nodes $u$ and $v$ with respect to some node cost function (we discuss below the specific node cost functions that we use).
Then, given the aligned clusters $a_i$, $i=1,\ldots,n$, the node
conservation measure is $S_N = \frac{1}{n}
\sum_{i=1}^n{ \frac{1}{{\abs{a_i} \choose 2}} \sum_{(u,v) \in
\mathcal{P}(a_i)} s(u,v) },$ where $\abs{a_i}$ is the size of $a_i$,
and $\mathcal{P}(a_i)$ is the set of all pairs of nodes in $a_i$.
Since multiMAGNA++ maximizes a convex combination of edge and node
conservation measures, the values of $S_N$ should ideally be in the
same range as the values of $S_E$.  Given that CIQ score lies between
0 to 1, in order to let $S_N$ also lie in that range, each $s(u,v)$
needs to lie between 0 and 1.  MultiMAGNA++, like MAGNA++, allows for
using any kind of node cost function to compute node conservation.

An NA method can use network topology alone in the alignment
construction process, or it can also include biological information
external to network topology, such as sequence information, while
constructing alignments.  We study the effect on alignment quality
when constructing alignments using only network topology versus also
including sequence information.
When constructing topology-only alignments, we let multiMAGNA++ optimize a combination of CIQ (corresponding to edge conservation $S_E$) and graphlet degree vector similarity (GDVS) (corresponding to $s(u,v)$ that is needed to compute node conservation $S_N$), where GDVS is a sensitive measure of topological similarity of extended neighborhoods of two nodes \citep{Milenkovic2008}. To combine $S_E$ and $S_N$, we use $\alpha=0.5$, to give equal contribution to edge conservation and node conservation.
When constructing topology+sequence alignments, we let multiMAGNA++ optimize a convex combination of: 1) the fitness function that we optimize for topology-only alignments (i.e., $0.5 \times \mbox{CIQ} + 0.5 \times \mbox{GDVS}$) and 2) BLAST sequence similarity as captured by E-value  \citep{BLAST}, as follows: $0.5 \times (0.5 \times \mbox{CIQ} + 0.5 \times \mbox{GDVS}) + 0.5 \times \mbox{E-value}$, where E-value is a commonly used node cost function for protein similarity \citep{IsoRankN,BEAMS,FUSE}. This way, when constructing topology+sequence alignments, we give equal contribution to the topological part (i.e., CIQ and GDVS combined) and the sequence part (i.e., E-value) of the fitness function.
Here, GDVS already lies in the [0,1] range, as required, but E-value does not. So, we convert E-value into a measure of similarity that lies between 0 and 1 (Section \ref{sec:supplfitness}), before we combine this measure with GDVS.

\subsubsection{Tying the GA together}
\label{sec:params}

We have discussed the components of our multiMAGNA++ that are needed
to optimize the above fitness function using a population of MNAs.
Additional parameters are: 1) how to generate the initial population;
2) which population size to use; 3) how to choose which individuals of
the population to cross; and 4) how many generations to run the
algorithm for.  For our choices of these parameters' values, see
Section
\ref{sec:supplparams}; we rely on our comprehensive
evaluation of the optimal parameter values conducted in our previous
MAGNA++ work.

\subsubsection{Running time and complexity}
\label{sec:complexity}

MAGNA++ and multiMAGNA++ evolve a population of $p$ alignments over $N$ generations.
For every generation, the methods perform two significant computations: (1) alignment quality for $p$ alignments with respect to both node and edge conservation, and (2) crossover of $O(p)$ pairs of parent alignments.
Given two networks $G_1(V_1,E_1)$ and $G_2(V_2,E_2)$, where $|V_1| \leq |V_2|$, MAGNA++ has a time
complexity of $O(N p |V_2| + N p (|E_1|+|E_2|))$, since computing node conservation takes $O(|V_2|)$, computing edge conservation takes $O(|E_1|+|E_2|)$, and computing crossover takes $O(|V_2|)$.
MAGNA++ improved upon the time complexity of MAGNA \citep{MAGNA++}, and MAGNA improved upon the complexity of existing (PNA) methods \citep{MAGNA}.
Hence, it is likely that multiMAGNA++ will be faster than existing MNA methods, most of which are also MNA-equivalents of their PNA versions, just as multiMAGNA++ is an MNA-equivalent of MAGNA++.
In particular, let $G_1(V_1,E_1), \ldots, G_k(V_k,E_k)$ be $k$ networks ordered by size as above, where $|E| =
\sum_{l=1}^k |E_l|$.
Then, the time complexity of multiMAGNA++ is $O(N p k |V_k| + N p
|E|)$, since computing node conservation takes $O(k|V_k|)$, computing
edge conservation takes $O(|E|)$, and computing crossover takes
$O(k|V_k|)$.  Clearly, the time complexity of multiMAGNA++ scales
linearly with the total number of nodes and edges.  Also, unlike with
the existing MNA methods, the complexity of multiMAGNA++ scales
linearly (rather than exponentially in, for example, GEDEVO-M's case)
with the number of networks.  Importantly, since calculating alignment
quality tends to be a bottleneck for multiMAGNA++, and since alignment
quality can be calculated independently for each alignment, we
parallelize this calculation in order to achieve a further
speedup. For multiMAGNA++, this results in speedup that is almost
linear in terms of the number of CPU cores used.

\subsection{Evaluation}
\label{sec:eval}

\subsubsection{Dataset}
\label{sec:dataset}

We use five PPI network sets: a network set with known true node mapping and four network sets with unknown node mapping.

{\em Networks with known node mapping.}  This network set, Yeast+\%LC,
has been used by many existing studies
\citep{GRAAL,HGRAAL,MIGRAAL,GHOST,MAGNA}.  It contains a
high-confidence {\em S. cerevisiae} (yeast) PPI network with 1,004
proteins and 8,323 PPIs \citep{YeastLC}, along with five yeast
lower-confidence networks that add PPIs of decreasing confidence to
the high-confidence network.  We align all six networks at once.  We
know the true node mapping since the networks contain the same nodes.
Thus, we can evaluate how accurately each MNA method reconstructs this
mapping (Section
\ref{sec:topquality}).

{\em Networks with unknown node mapping.}  The four network sets with
unknown node mapping are PHY1, PHY2, Y2H1 and Y2H2
\citep{LocalVsGlobal}.  Each network set contains PPI data of four
species, {\em S. cerevisiae} (yeast/Y), {\em D. melanogaster} (fly/F),
{\em C. elegans} (worm/W), and {\em H. sapiens} (human/H), obtained
from BioGRID
\citep{BioGRID} in November 2014.  The networks in the four sets were extracted
based on the following interaction types and confidence levels of the
PPIs: (i) all physical PPIs supported by at least one publication
(PHY1), (ii) all physical PPIs supported by at least two publications
(PHY2), (iii) only yeast two-hybrid physical PPIs supported by at
least one publication (Y2H1), and (iv) only yeast two-hybrid physical
PPIs supported by at least two publications (Y2H2).  Just as in our
recent work \citep{LocalVsGlobal}, we use network sets with different
PPI types (all physical vs. only yeast two-hybrid) to test the
robustness of our approach to the choice of PPI data type.  We use
network sets with PPIs supported by at least two publications since
those PPIs are believed to be more reliable than PPIs supported by
only one publication \citep{LitCurPPIdatasets}.  For each of the
networks, we use only its largest connected component (Table \ref{tab:nets}).  The largest connected component of the fly and
worm networks in the PHY2 and Y2H2 sets are too small (53-331 nodes)
for analyses.  Thus, we remove the fly and worm networks from both
PHY2 and Y2H2, resulting in each of the two sets containing only two
networks.  Since we cannot measure how accurately the aligners
reconstruct the true node mapping when aligning networks with unknown
node mapping, we use alternative alignment accuracy measures (Section
\ref{sec:bioquality}).

Some of these alternative measures rely on Gene Ontology (GO)
annotations of proteins \citep{GOdata}.
We use GO data obtained from the Gene Ontology database in January 2016.
We use only GO annotations that were obtained experimentally.

\subsubsection{Existing methods we evaluate against}
\label{sec:othermethods}

The methods we evaluate against are all existing MNA methods for which the code is available and could be run without errors. Namely, we comprehensively evaluate against:
i) IsoRankN \citep{IsoRankN}, ii)
MI-Iso \citep{Faisal2014}, iii) GEDEVO-M \citep{GEDEVOM}, iv)
BEAMS \citep{BEAMS}, and FUSE \citep{FUSE}.  Also, we evaluate our approach and each of the
existing approaches against their corresponding random MNA
counterparts, to ensure statistical significance of each result (see
below).  
We tried to evaluate against SMETANA \citep{SMETANA}, but were unable to do so due to SMETANA's high memory usage and high runtime on the larger network sets.
Similarly, we tried to evaluate against NetCoffee \citep{NetCoffee}, but we were unable to do so on our machines, because of incompatibility of NetCoffee's library dependencies.
Finally, we do not evaluate against the remaining existing method, CSRW \citep{RandomWalk}, since its publication does not make available the code for this method.

To fairly evaluate the methods, we study the effect on alignment
quality of (i) using only network topology while constructing
alignments (resulting in topology-only alignments) versus (ii) also
including sequence information into the alignment construction process
(resulting in topology+sequence alignments).  For topology-only
alignments, we set method parameters to ignore any sequence
information (Table \ref{tab:params}).  All methods
except BEAMS and FUSE can be run in the topology-only mode.  For
topology+sequence alignments, we set method parameters to include
BLAST sequence information (Section
\ref{sec:fitness} and Table \ref{tab:params}).  All
methods but GEDEVO-M can be run in the topology+sequence mode.

The MNA methods that we evaluate are classified into one-to-one (1-1)
and many-to-many (m-m) aligners.  Of the methods, GEDEVO-M, FUSE, and
multiMAGNA++ are 1-1 aligners while IsoRankN, MI-Iso, and BEAMS are
m-m aligners.  Since 1-1 and m-m aligners result in different outputs,
meaning that the aligned clusters produced by 1-1 aligners contain at
most one node from each network while m-m aligners have no such
restriction, it is more fair to compare 1-1 aligners with other 1-1
aligners, and m-m aligners with other m-m aligners.  Comparison of 1-1
aligners with m-m aligners needs to be taken with caution due to their
different output types.  Yet, we include such comparison, since only
two of the existing methods (GEDEVO-M and FUSE) are 1-1, i.e., is of the same
type as our proposed multiMAGNA++ approach, and we want to include
more MNA approaches into the comparison to properly demonstrate the
superiority of our approach.

To allow for as fair as possible comparison of 1-1 and m-m aligners,
we compute the statistical significance of each approach's alignment
quality score, in order to compare the resulting $p$-values between
the different MNA approaches instead of (or at least in addition to)
comparing the approaches' raw alignment quality scores. Namely, for
each approach and each of its alignments (depending on the input
networks), we construct a set of 10,000 corresponding random
alignments (10,000 is what was practically possible given the
relatively large running time of computing the functional alignment
quality scores), under a null model that accounts for characteristics
of both the given approach and the input networks. That is, each
random alignment conserves the number of clusters and the cluster size
distribution of the corresponding actual alignment. Then, we compute
the $p$-value of the given alignment quality score as
the frequency of obtaining equal or better score among the 10,000
random alignments. This way, by comparing $p$-values of the
different approaches instead of (or in addition to, in case $p$-values
of the different approaches are tied) the approaches' raw scores,
where the $p$-values account for the null model of each approach, we
are aiming to account for the differences between output types of 1-1
and m-m MNA approaches, in order to allow for their more fair
comparison.  We consider an alignment score to be significant if its $p$-value is
less than 0.001.  Note that we obtain qualitatively identical results
when we use a more flexible $p$-value threshold of 0.01.

\subsubsection{Topological alignment quality measures}
\label{sec:topquality}

We propose three new measures of topological alignment quality for MNA, as described below.
Whenever we calculate alignment quality (here or in Section \ref{sec:bioquality}), we only consider aligned clusters with at least two nodes, unless specified otherwise.

\vspace{0.5em}

\noindent
{\bf Adjusted node correctness (NCV-MNC).}  A good MNA approach should
find aligned clusters that are internally consistent with respect to
protein labels.  For networks with known node mapping, labels
correspond to protein names. In this case, we use an existing notion
of node correctness (MNC, defined below) to measure internal cluster
consistency, and we consider this to be a topological alignment
quality measure.  First, we use an existing notion of normalized
entropy (NE) to measure how likely it is to observe in a given
cluster, at random, the same or higher level of internal consistency
with respect to protein names.  Given cluster $c$, $\textrm{NE}(c) =
-\frac{1}{\log d} \sum_{i=1}^d{p_i \log p_i}$, where $d$ is the number
of unique protein names in $c$, and $p_i$ is the fraction of nodes in
$c$ with protein name $i$.  The lower the NE, the more consistent the
cluster.  Then, we let MNC be one minus the mean of NEs across all
clusters in the alignment.  For networks with unknown node mapping,
protein labels correspond to their GO terms. In this case, since GO
terms capture proteins' functional information, we consider the
corresponding measure of internal cluster consistency to be a
functional alignment quality measure, and we describe this measure in
more detail in Section \ref{sec:bioquality}.

A good MNA approach should also align (or cover) many of the proteins
from the aligned networks.  So, we combine the above notion of cluster
consistency (i.e. MNC) with an existing notion of node coverage (NCV,
defined below) into a new measure that we call {\em adjusted node
correctness}.  NCV is the fraction of all nodes that are part of the
alignment (i.e., of the aligned clusters with two or more nodes) out
of all nodes in the networks.  Then, we define adjusted node
correctness as: $\textrm{NCV-MNC} =
\sqrt{(\textrm{NCV})(\textrm{MNC})}$.  This geometric mean penalizes
alignments that have a low alignment quality score with respect to at
least one of NCV or MNC.

\vspace{0.5em}

\noindent
{\bf Adjusted cluster interaction quality (NCV-CIQ).} Further, a good
MNA approach should find a large amount of network structure that is
common to many (ideally all) of the aligned networks, which is
typically referred to as edge conservation. Here, we rely on the
existing CIQ measure, which we have already defined in Section
\ref{sec:fitness}.  CIQ can be seen as a generalization of the
established S$^3$ \citep{MAGNA} edge conservation measure from PNA to
MNA.  Just as above, since we want the conserved edges to cover many
nodes, we combine CIQ with NCV into a new measure, {\em adjusted
cluster interaction quality}, as follows: $\textrm{NCV-CIQ} =
\sqrt{(\textrm{NCV})(\textrm{CIQ})}$.

\vspace{0.5em}

\noindent
{\bf Largest common connected subgraph (LCCS).} Finally, a good MNA
approach should group the aligned edges to form connected and dense
network regions (as opposed to the aligned edges being isolated in a
random manner).  To capture this, we rely on an established notion of
the size of LCCS, but we propose a novel generalization of this notion
from PNA to MNA.  The details of this novel LCCS measure are as
follows.

Given an MNA of $k$ networks $G_1(V_1,E_1)$, $G_2(V_2,E_2)$, $\ldots$,
$G_k(V_k,E_k)$, we define the fully conserved common subgraph as the
graph in which each aligned cluster is fused into a supernode and
there is an edge between two supernodes if and only if there is an
edge belonging to each of the $k$ networks that connects the two
aligned clusters.  Then, the LCCS is the largest connected component
of the fully conserved common subgraph.  To measure the quality of the
LCCS of the given alignment by simultaneously accounting for the
LCCS's number of nodes as well as its number of edges, we extend the
PNA-based LCCS measure introduced in \citep{MAGNA} to MNA.  Let $n$ be
the number of nodes in the LCCS and $n_{max} =
\min\{|V_1|,\ldots,|V_k|\}$ be the maximum possible number of nodes in
the LCCS.  Let $e$ be the number of edges in the LCCS and $e_{max} =
\min\{|E_1(\textrm{LCCS})|,\ldots,|E_k(\textrm{LCCS})|\}$ be the
maximum possible number of edges in the LCCS, where
$E_l(\textrm{LCCS})$ is the set of edges induced by the network nodes
in the LCCS on network $G_l$.  Then, $\textrm{LCCS} =
\sqrt{\frac{n}{n_{max}} \frac{e}{e_{max}}}$; this score is high
when the LCCS both has many nodes and is dense.

\subsubsection{Functional alignment quality measures}
\label{sec:bioquality}

We use three existing measures of functional alignment quality for
MNA, modifying in the process some measures to allow for MNA instead of
just PNA, as described below.  These measures rely on GO data. Since
many GO annotations are obtained via sequence analyses, and since we
also use sequence information in the alignment construction process,
to avoid a circular argument, we only use GO annotations that have
been obtained experimentally.

\vspace{0.5em}

\noindent
{\bf Mean Normalized Entropy (MNE).}  Recall from Section
\ref{sec:topquality} that a good MNA should have clusters that are
internally consistent.  When the true node mapping is unknown, we use
GO terms to measure internal consistency.  For this purpose, we use
the same NE measure as in Section \ref{sec:topquality}, where now $d$
is the number of unique GO terms, and $p_i$ is the ratio of the number
of proteins annotated with GO term $i$ to the total number of
protein-GO term annotations (independent of the GO term) in the
cluster.  The lower the NE, the more consistent the cluster is with
respect to the GO terms.  We measure the internal consistency over all
clusters using MNE, the mean of the NEs across all clusters in the
alignment \citep{IsoRankN}.

\vspace{0.5em}

\noindent
{\bf GO correctness (GC).}  We extend the existing notion of GO
correctness from PNA to MNA as another measure of internal cluster
consistency \citep{GRAAL}.  GO correctness measures the extent to
which pairs of proteins that are aligned together are annotated with
the same GO term(s).  For MNA, we consider two proteins to be aligned
together if they are in the same cluster.  Formally, to calculate GC,
we first transform an MNA consisting of aligned node clusters into a
list of aligned node pairs.  This is done by populating this list with
all pairs of proteins that are in the same aligned cluster.  Then, we
filter this list to keep only pairs in which each of the two proteins
has at least one GO term.  Given the resulting filtered list, GC is
the fraction of the filtered protein pairs in which the two proteins
share at least one GO term.  In this analysis, we ignore GO terms that
are associated with only one protein in the MNA.

\vspace{0.5em}

\noindent
{\bf Accuracy of protein function prediction.}  We extend the protein
function prediction approach by \citet{LocalVsGlobal} from PNA to MNA.
In a leave-one-out-cross-validation-like manner, we measure how well
we can predict function (i.e., GO term) of a given protein based on
the protein's aligned partner when we hide the protein's functional
information, repeating this for each currently annotated protein from
the above filtered list of aligned protein pairs and each of its GO
terms.  Then, given all predicted protein-GO term prediction
associations, we calculate accuracy of the predictions via precision,
recall, and F-score measures.  Formally, let $X$ be the set of
predicted protein-GO term associations, and let $Y$ be the set of true
protein-GO term associations.  Then, the precision of protein function
prediction is: P-PF $ = \frac{\abs{X \cap Y}}{\abs{X}}$. The recall
is: R-PF $ = \frac{\abs{X \cap Y}}{\abs{Y}}$. The F-score, F-PF, is
the harmonic mean of precision and recall.  In this analysis, we
ignore GO terms that are associated with only one protein in the MNA.


\section{Results and Discussion}
\label{sec:results}

We compare multiMAGNA++ (Section \ref{sec:magna}) to five existing MNA
methods (IsoRankN, MI-Iso, GEDEVO-M, BEAMS, FUSE; Section \ref{sec:othermethods}) on a network set with known
node mapping, Yeast+\%LC, and four network sets with unknown node
mapping, PHY1, PHY2, Y2H1, and Y2H2 (Section \ref{sec:dataset}). 
We measure alignment quality using both topological and functional alignment quality measures (Sections \ref{sec:topquality} and \ref{sec:bioquality}). We measure the statistical significance of each result (Section \ref{sec:othermethods}).

Recall that we consider topology-only alignments as well as topology+sequence alignments (Section \ref{sec:othermethods}). For each data set and alignment quality measure, since we want to give each method the best case advantage, we do the following. Henceforth, when we report results for the given method, we do so for the best of its topology-only alignment and its topology+sequence alignment, unless otherwise noted. Note that even if we restrict our method evaluation only to topology-only alignments, or only to topology+sequence alignments, the results are qualitatively similar, and all detailed results are reported in the Supplement.

\subsection{Method comparison in terms of accuracy}

\subsubsection{Network set with known node mapping}
\label{sec:knownmap}

\begin{table}[h]
\centering
\scriptsize
\begin{tabular}{|p{22mm}|p{18mm}|p{18mm}|p{18mm}|p{18mm}|p{18mm}|p{18mm}|} \hline
 & \multicolumn{3}{c|}{\scriptsize Topological measures} & \multicolumn{3}{c|}{\scriptsize Functional measures} \\ \hline
   \scriptsize{ Method   }
 & \scriptsize{ NCV-MNC  }
 & \scriptsize{ NCV-CIQ  }
 & \scriptsize{ LCCS     }
 & \scriptsize{ MNE      }
 & \scriptsize{ GC       }
 & \scriptsize{ F-score  }
 \\ \hline
 MI-Iso
 & 0.5638 \newline p $<$ 1$\text{e-}$4
 & 0.7908 \newline p $<$ 1$\text{e-}$4
 & 0.2508 \newline p $<$ 1$\text{e-}$4
 & 0.9508 \newline p $<$ 1$\text{e-}$4
 & 0.6137 \newline p $<$ 1$\text{e-}$4
 & 0.4709 \newline p $<$ 1$\text{e-}$4
 \\ \cline{1-7}
 IsoRankN
 & 0.8967 \newline p $<$ 1$\text{e-}$4
 & 0.9219 \newline p $<$ 1$\text{e-}$4
 & 0.8929 \newline p $<$ 1$\text{e-}$4
 & 0.9532 \newline p $<$ 1$\text{e-}$4
 & 0.9748 \newline p $<$ 1$\text{e-}$4
 & 0.9367 \newline p $<$ 1$\text{e-}$4
 \\ \cline{1-7}
 BEAMS
 & 0.8215 \newline p $<$ 1$\text{e-}$4
 & 0.8342 \newline p $<$ 1$\text{e-}$4
 & 0.7827 \newline p $<$ 1$\text{e-}$4
 & 0.9667 \newline p $<$ 1$\text{e-}$4
 & 0.9882 \newline p $<$ 1$\text{e-}$4
 & 0.8625 \newline p $<$ 1$\text{e-}$4
 \\ \cline{1-7}
 FUSE
 & 0.6299 \newline p $<$ 1$\text{e-}$4
 & 0.6943 \newline p $<$ 1$\text{e-}$4
 & 0.5123 \newline p $<$ 1$\text{e-}$4
 & 0.9618 \newline p $<$ 1$\text{e-}$4
 & 0.7744 \newline p $<$ 1$\text{e-}$4
 & 0.5953 \newline p $<$ 1$\text{e-}$4
 \\ \cline{1-7}
 GEDEVO-M
 & 0.2969 \newline p $<$ 1$\text{e-}$4
 & 0.8574 \newline p $<$ 1$\text{e-}$4
 & 0.4104 \newline p $<$ 1$\text{e-}$4
 & 0.9610 \newline p $<$ 1$\text{e-}$4
 & 0.4333 \newline p $<$ 1$\text{e-}$4
 & 0.1586 \newline p $<$ 1$\text{e-}$4
 \\ \cline{1-7}
 \scripty{multiMAGNA++}
 & {\bf 0.9241 \newline p $<$ 1$\textbf{e-}$4 }
 & {\bf 0.9574 \newline p $<$ 1$\textbf{e-}$4 }
 & {\bf 0.9201 \newline p $<$ 1$\textbf{e-}$4 }
 & {\bf 0.9341 \newline p $<$ 1$\textbf{e-}$4 }
 & {\bf 0.9897 \newline p $<$ 1$\textbf{e-}$4 }
 & {\bf 0.9392 \newline p $<$ 1$\textbf{e-}$4 }
 \\ \hline
\end{tabular}
\normalsize
\caption{Alignment accuracy of different MNA approaches for the Yeast+\%LC network set in terms of topological NCV-MNC, NCV-CIQ, and LCCS measures and functional MNE, GO correctness (GC), and F-score measures. The symbol ``p'' signifies $p$-values of the observed alignment scores. For each alignment quality measure, and for each network set, the best method (i.e., the method with the lowest $p$-value, or the method with the best alignment quality score if the $p$-values are tied) is bolded. The alignment scores that are not statistically significant, if any, are greyed out and italicized. Note that for MNE, the lower the score, the better the alignment quality. For all other measures, the higher the score, the better the alignment quality. For equivalent results broken down by topology-only alignments and by topology+sequence alignments, see Tables \ref{tab:all_yeastlc_top}-\ref{tab:all_yeastlc_topseq} and Figures \ref{fig:mnc_yeastlc}-\ref{fig:fscore}.}
\label{tab:all_yeastlc_best}
\end{table}

While on this data all methods produce statistically significant alignments with respect to all alignment quality measures, multiMAGNA++ outperforms all existing methods with respect to all topological and functional alignment quality measures (Table \ref{tab:all_yeastlc_best}). 
Furthermore, only multiMAGNA++, IsoRankN, and BEAMS perform well consistently for each of the measures.
Note that GEDEVO-M, which optimizes edge conservation like multiMAGNA++, does not perform as well as the other methods.
(In the original GEDEVO-M publication \citep{GEDEVOM}, nearly perfect alignments of the networks with known node mapping were reported with respect to node conservation. We found that this is because GEDEVO-M's implementation uses the lexicographic ordering of the node names. After we remove this bias by renaming the node names to randomly generated strings, the accuracy of GEDEVO-M drops, as reported in our results.)
The inferior behavior of GEDEVO-M compared to the other methods could be because GEDEVO-M can use only topological information when constructing alignments, while all other methods can use both topological and sequence information. Namely, Yeast+\%LC, comprising six networks that all have the
same set of nodes, contains many inter-network pairs of nodes that are
the same proteins (this does not happen in networks with unknown node
mapping that have different node sets). For such pairs, sequence
similarity scores are significantly higher than for any other pairs of
nodes.  These inter-network node pairs can potentially form aligned
clusters that have very high intra-cluster sequence similarity due to
the node pairs being the same protein.
This may be the main reason
why on Yeast+\%LC, the MNA approaches perform much better for topology+sequence alignments compared to topology-only alignments (Tables \ref{tab:all_yeastlc_top} and \ref{tab:all_yeastlc_topseq}).
This is not the case for the network
sets with unknown node mapping, PHY1, PHY2, Y2H1, and Y2H2, as we
see in the following section.


\subsubsection{Networks with unknown node mapping}
\label{sec:unknownmap}

\begin{table}[h!]
\begin{center}
\scriptsize
\begin{tabular}{|l|p{22mm}|p{18mm}|p{18mm}|p{18mm}|p{18mm}|p{18mm}|} \hline
 & & \multicolumn{2}{c|}{\scriptsize Topological measures} & \multicolumn{3}{c|}{\scriptsize Functional measures} \\ \hline
 & \scriptsize{Method   }
 & \scriptsize{NCV-CIQ  }
 & \scriptsize{LCCS     }
 & \scriptsize{MNE      }
 & \scriptsize{GC       }
 & \scriptsize{F-score  }
 \\ \hline
\parbox[t]{2mm}{\multirow{12}{*}{\rotatebox[origin=c]{90}{PHY1}}}
 & MI-Iso
 & 0.2517 \newline p $<$ 1$\text{e-}$4
 & {\it \textcolor{gray!125} { 0.0000 \newline p = 1.000 }}
 & 0.8224 \newline p $<$ 1$\text{e-}$4
 & 0.2732 \newline p $<$ 1$\text{e-}$4
 & 0.0492 \newline p $<$ 1$\text{e-}$4
 \\ \cline{2-7}
 & IsoRankN
 & 0.1012 \newline p $<$ 1$\text{e-}$4
 & 0.0258 \newline p $<$ 1$\text{e-}$4
 & {\bf 0.7977 \newline p $<$1$\textbf{e-}$4 }
 & {\bf 0.3279 \newline p $<$1$\textbf{e-}$4 }
 & {\bf 0.0838 \newline p $<$1$\textbf{e-}$4 }
 \\ \cline{2-7}
 & BEAMS
 & 0.3250 \newline p $<$ 1$\text{e-}$4
 & {\it \textcolor{gray!125} { 0.0000 \newline p = 1.000 }}
 & {\it \textcolor{gray!125} { 0.8944 \newline p = 0.895 }}
 & {\it \textcolor{gray!125} { 0.4084 \newline p = 1.000 }}
 & 0.0457 \newline p $<$ 1$\text{e-}$4
 \\ \cline{2-7}
 & FUSE
 & 0.0679 \newline p = 5\text{e-}4
 & {\it \textcolor{gray!125} { 0.0000 \newline p = 1.000 }}
 & 0.8781 \newline p $<$ 1$\text{e-}$4
 & 0.2268 \newline p $<$ 1$\text{e-}$4
 & 0.0472 \newline p $<$ 1$\text{e-}$4
 \\ \cline{2-7}
 & GEDEVO-M
 & 0.3554 \newline p $<$ 1$\text{e-}$4
 & {\bf 0.1613 \newline p $<$1$\textbf{e-}$4 }
 & {\it \textcolor{gray!125} { 0.9205 \newline p = 1.000 }}
 & 0.1721 \newline p $<$ 1$\text{e-}$4
 & 0.0324 \newline p $<$ 1$\text{e-}$4
 \\ \cline{2-7}
 & multiMAGNA++
 & {\bf 0.4046 \newline p $<$1$\textbf{e-}$4 }
 & 0.1064 \newline p $<$ 1$\text{e-}$4
 & 0.8449 \newline p $<$ 1$\text{e-}$4
 & 0.1759 \newline p $<$ 1$\text{e-}$4
 & 0.0353 \newline p $<$ 1$\text{e-}$4
 \\ \hline
\parbox[t]{2mm}{\multirow{12}{*}{\rotatebox[origin=c]{90}{Y2H1}}}
 & MI-Iso
 & 0.1935 \newline p $<$ 1$\text{e-}$4
 & 0.0264 \newline p $<$ 1$\text{e-}$4
 & {\it \textcolor{gray!125} { 0.8992 \newline p = 0.396 }}
 & 0.2092 \newline p $<$ 1$\text{e-}$4
 & 0.0382 \newline p $<$ 1$\text{e-}$4
 \\ \cline{2-7}
 & IsoRankN
 & 0.1315 \newline p $<$ 1$\text{e-}$4
 & 0.0264 \newline p $<$ 1$\text{e-}$4
 & {\bf 0.8447 \newline p $<$1$\textbf{e-}$4 }
 & 0.3247 \newline p $<$ 1$\text{e-}$4
 & 0.0822 \newline p $<$ 1$\text{e-}$4
 \\ \cline{2-7}
 & BEAMS
 & 0.2856 \newline p $<$ 1$\text{e-}$4
 & {\it \textcolor{gray!125} { 0.0000 \newline p = 1.000 }}
 & {\it \textcolor{gray!125} { 0.9159 \newline p = 0.363 }}
 & {\bf 0.3945 \newline p $<$1$\textbf{e-}$4 }
 & {\bf 0.0856 \newline p $<$1$\textbf{e-}$4 }
 \\ \cline{2-7}
 & FUSE
 & 0.0480 \newline p $<$ 1$\text{e-}$4
 & {\it \textcolor{gray!125} { 0.0000 \newline p = 1.000 }}
 & 0.8781 \newline p $<$ 1$\text{e-}$4
 & 0.2369 \newline p $<$ 1$\text{e-}$4
 & 0.0483 \newline p $<$ 1$\text{e-}$4
 \\ \cline{2-7}
 & GEDEVO-M
 & 0.4511 \newline p $<$ 1$\text{e-}$4
 & 0.0722 \newline p $<$ 1$\text{e-}$4
 & {\it \textcolor{gray!125} { 0.9032 \newline p = 0.919 }}
 & 0.1879 \newline p = 6\text{e-}4
 & 0.0347 \newline p $<$ 1$\text{e-}$4
 \\ \cline{2-7}
 & multiMAGNA++
 & {\bf 0.4943 \newline p $<$1$\textbf{e-}$4 }
 & {\bf 0.1088 \newline p $<$1$\textbf{e-}$4 }
 & 0.8899 \newline p = 6\text{e-}4
 & 0.2040 \newline p $<$ 1$\text{e-}$4
 & 0.0428 \newline p $<$ 1$\text{e-}$4
 \\ \hline
\end{tabular}
\normalsize
\end{center}
\caption{Alignment accuracy of different MNA approaches for the PHY1 and Y2H1 network sets in terms of topological NCV-CIQ and LCCS measures and functional MNE, GO correctness (GC), and F-score measures. The symbol ``p'' signifies $p$-values of the observed alignment scores. For each alignment quality measure, and for each network set, the best method (i.e., the method with the lowest $p$-value, or the method with the best alignment quality score if the $p$-values are tied) is bolded. The alignment scores that are not statistically significant, if any, are greyed out and italicized. Note that for MNE, the lower the score, the better the alignment quality. For all other measures, the higher the score, the better the alignment quality. Equivalent results for the remaining networks with unknown true node mapping (PHY2 and Y2H2) are shown in Table \ref{tab:all_all_best}. For equivalent results broken down by topology-only alignments and by topology+sequence alignments, see Tables \ref{tab:all_all_top}-\ref{tab:all_all_topseq} and Figures \ref{fig:ciq}-\ref{fig:fscore}.}
\label{tab:all_all_best_short}
\end{table}
Over all four network sets with unknown node mapping and all five alignment quality measures, multiMAGNA++ produces the most of statistically significant alignments. Namely, in only one of the $4 \times 5 = 20$ cases, its alignment is non-significant, while the existing methods have non-significant alignments in four to eight of the 20 cases (Table \ref{tab:all_all_best_short} and Table \ref{tab:all_all_best}).
This has important implications for predicting any new biological knowledge from a given method's alignment, since an alignment is practically meaningful only if it is statistically significant.

In terms of topological alignment quality, multiMAGNA++ is superior to MI-Iso, IsoRankN, and FUSE in all cases, and it is also superior to BEAMS and GEDEVO-M in 75\% of all cases (Table
\ref{tab:all_all_best_short} and Table \ref{tab:all_all_best}). 

In terms of functional alignment quality, multiMAGNA++ is overall comparable to all methods, with the exception of IsoRankN, which is the best performing method in most of the cases. However, IsoRankN's performance is data-specific, as it works great for some network sets but completely fails for others (such as PHY2; Table \ref{tab:all_all_best}), while multiMAGNA++ performs consistently well on all network sets.



\begin{figure}[h!]
\centering
\includegraphics[width=0.47\linewidth]{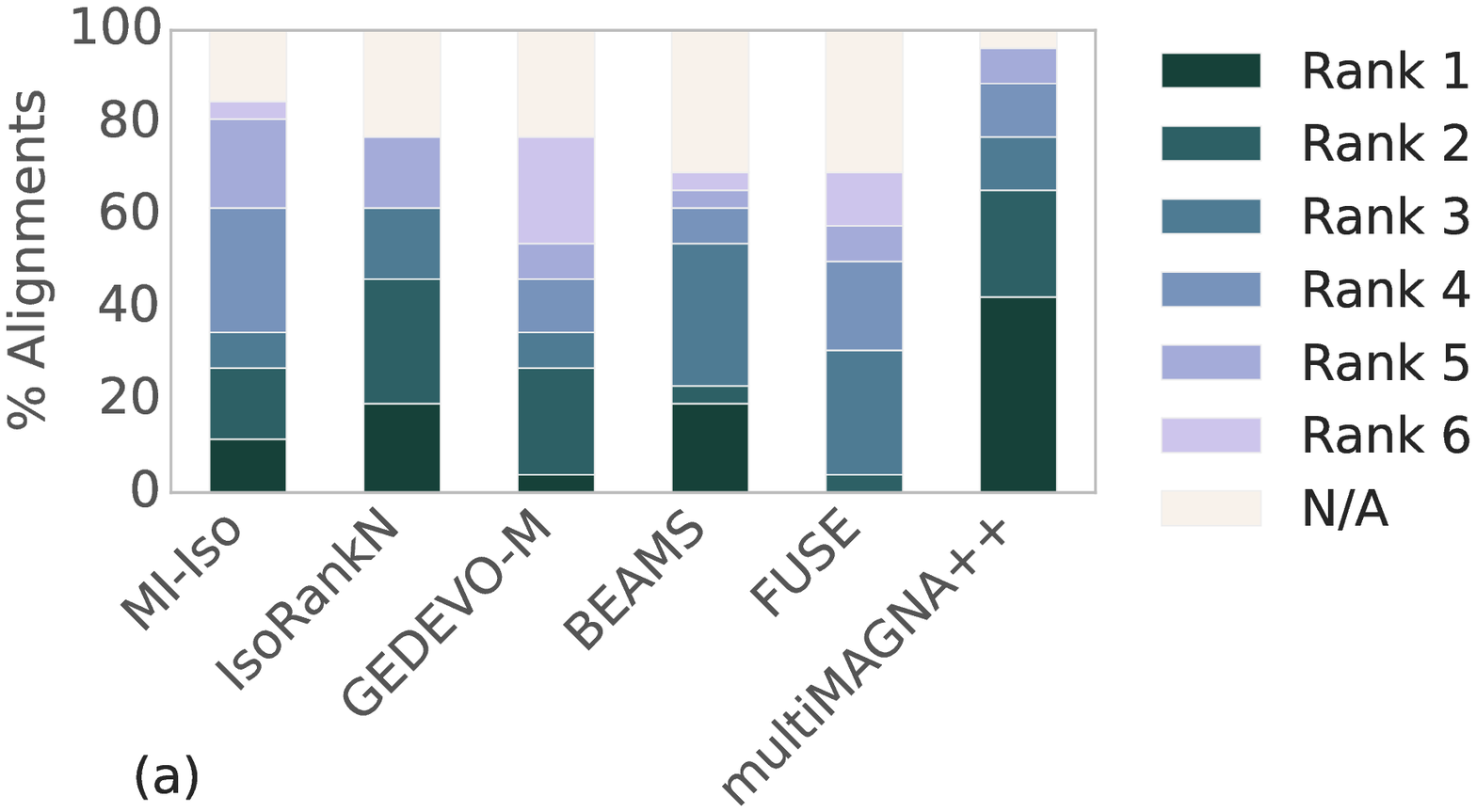}\\
\includegraphics[width=0.47\linewidth]{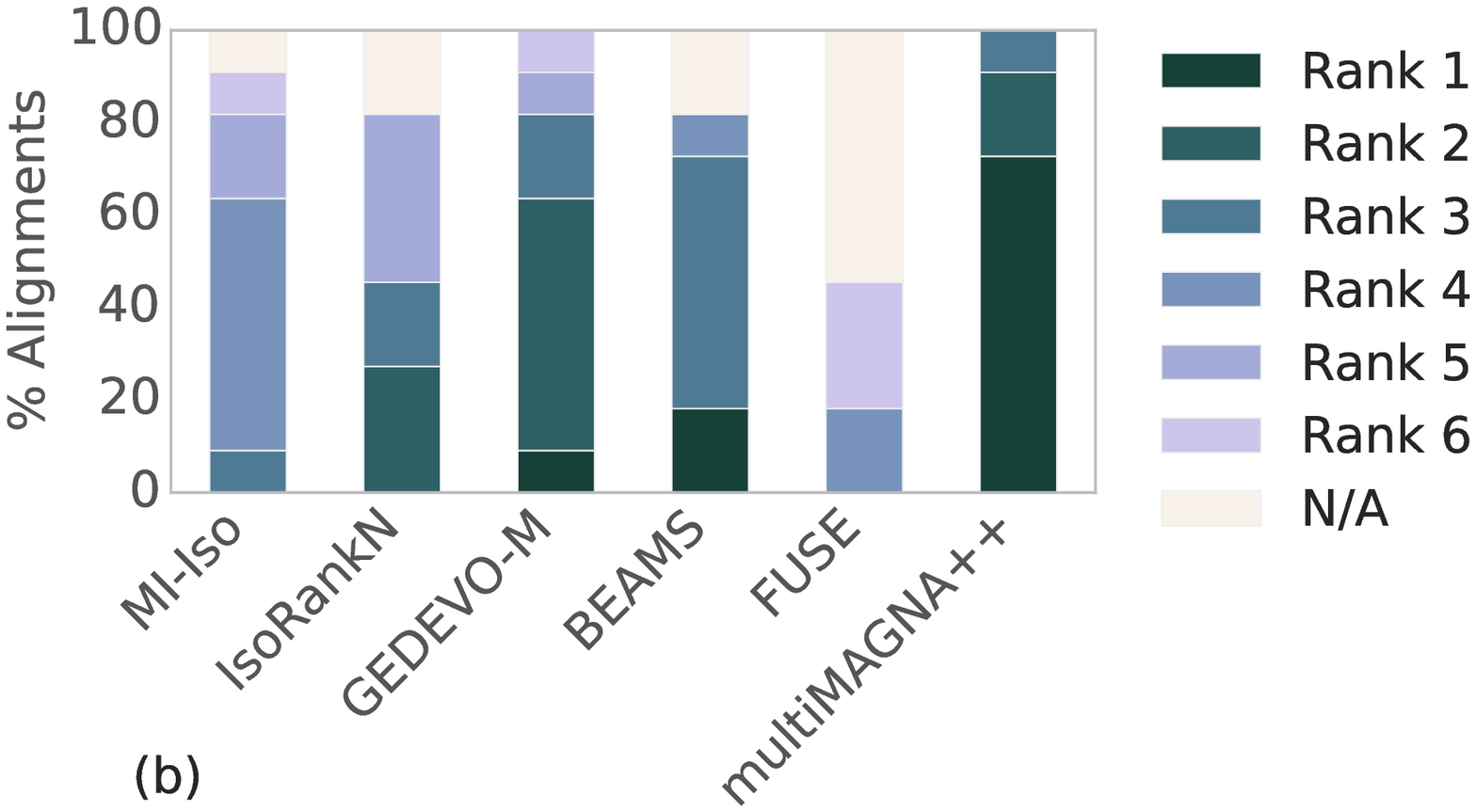}\\
\includegraphics[width=0.47\linewidth]{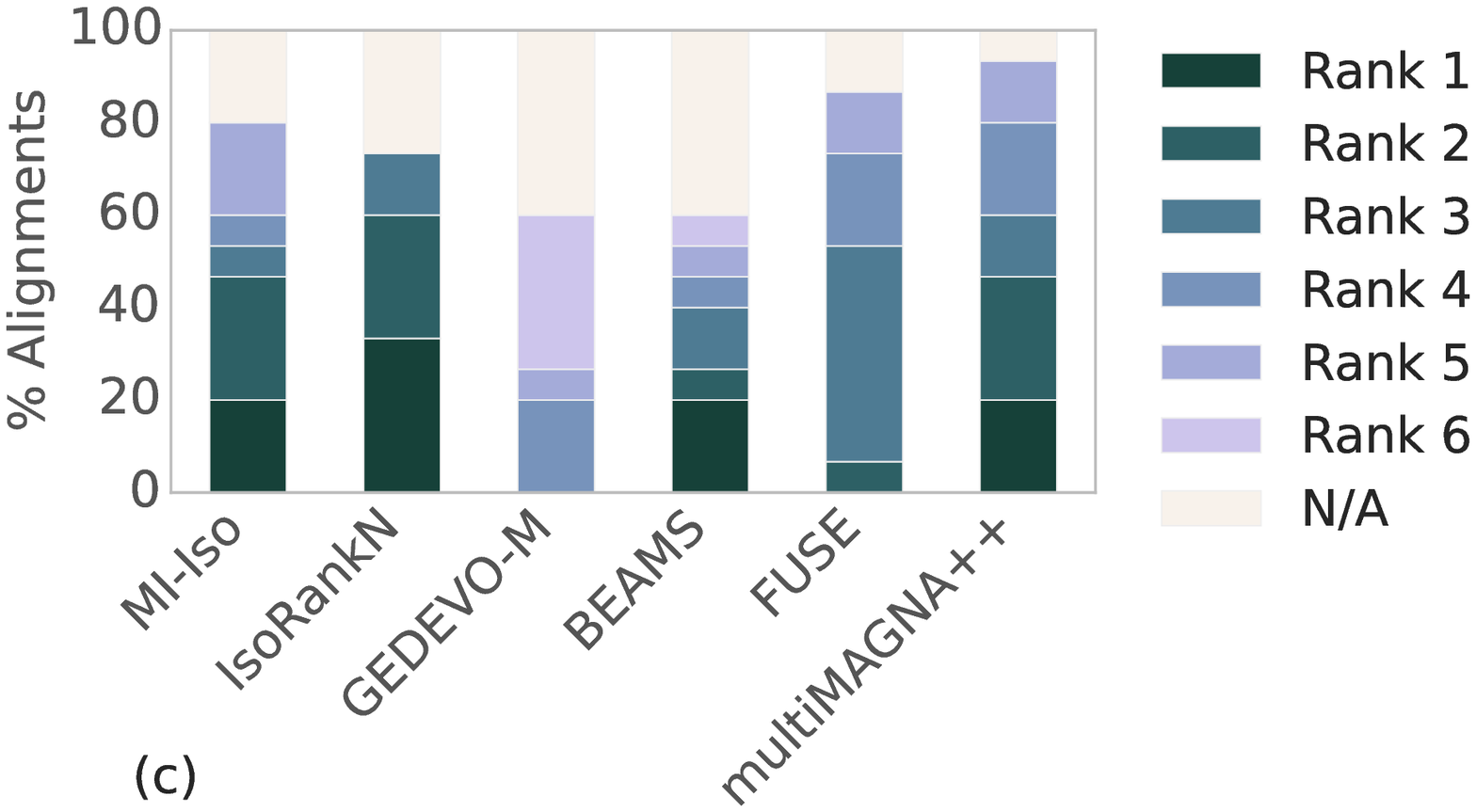}
\caption{Ranking of the MNA methods across all of Yeast+\%LC, PHY1, PHY2, Y2H1, and Y2H2 network sets with respect to \textbf{(a)} all measures, \textbf{(b)} topological  NCV-MNC, NCV-CIQ, and LCCS measures, and \textbf{(c)} functional MNE, GC, and F-score measures. The ranking of each method is expressed as a percentage of all evaluation tests in which the given method is the best performing (``Rank 1''), the second best performing (``Rank 2''), etc. aligner of all considered methods. By ``best'', we mean the method with the lowest $p$-value, or the method with the best alignment quality score if the $p$-values are tied. If an alignment score of a method is not statistically significant, the method is not ranked and is labelled as ``N/A''. For equivalent results broken down by topology-only alignments and by topology+sequence alignments, see Figure \ref{fig:all_rankings}.}
\label{fig:all_rankings_best}
\end{figure}



\subsubsection{Summary}

We summarize our results over all network sets and all alignment quality measures to identify the overall ranking of the different MNA methods, i.e., to identify the best (rank 1), second best (rank 2), third best (rank 3), etc. performing of all methods. MultiMAGNA++ has  barely any non-significant alignments (3.9\% of all cases), while the existing methods have 15.4\%--30.8\% non-significant alignments (Figure \ref{fig:all_rankings_best}(a)). This
makes multiMAGNA++ the best method in this context.
Further, multiMAGNA++ is superior to all other methods (rank 1) in more cases than any other method (Figure \ref{fig:all_rankings_best}(a)).
Importantly, only multiMAGNA++ and BEAMS are superior to all existing methods (i.e., have rank 1) in terms of both topological (Figure \ref{fig:all_rankings_best}(b)) and functional (Figure \ref{fig:all_rankings_best}(c)) alignment quality; all other methods have rank 1 with respect to at most one of topological and functional quality. In this context, multiMAGNA++ and BEAMS are outperforming the other methods, while at the same time, multiMAGNA++ beats BEAMS, especially in terms of topological alignment quality.
In conclusion, when comparing the different methods in terms of alignment accuracy, multiMAGNA++ drastically beats the existing methods with respect to topological alignment quality (Figure \ref{fig:all_rankings_best}(b)), and it is comparable to the existing methods with respect to functional alignment quality (Figure \ref{fig:all_rankings_best}(c)). 

\begin{figure}[h!]
\centering
\includegraphics[width=0.51\linewidth]{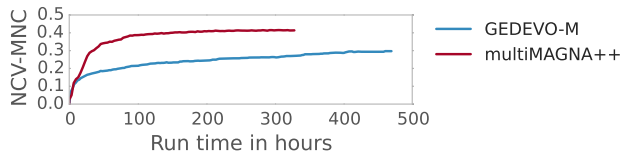}
\caption{NCV-MNC as a function of the number of hours spent by multiMAGNA++ and GEDEVO-M  when using a single thread, for topology-only alignments of Yeast+\%LC networks. Both multiMAGNA++ and GEDEVO-M are run for 100,000 generations. These are representative results. For equivalent results for the other  measures and for the networks with unknown node mapping, see Figures \ref{fig:time_ciq} and \ref{fig:time_lccs}.}
\label{fig:time_mnc}
\end{figure}


\begin{figure}[h!]
\centering
\includegraphics[width=0.45\linewidth]{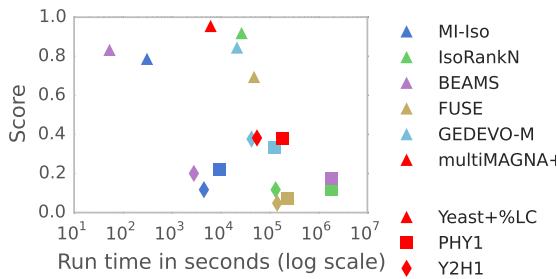}
\caption{NCV-CIQ as a function of time when using 64 threads for the three network sets with more than two networks (we leave out these results for PHY2 and Y2H2 that have two networks each; Section \ref{sec:dataset}). These are representative results. For equivalent results for the remaining measures, as well as for using a single thread, see Figures \ref{fig:timing_mnc}-\ref{fig:timing_lccs}.}
\label{fig:timing_ciq_best}
\end{figure}

\subsection{Method comparison in terms of  time complexity}
\label{sec:runtime}

First, we compare multiMAGNA++ with GEDEVO-M, as these methods are the
most similar since they are both evolutionary algorithms. This allows
us to observe their behavior over time to see how fast they converge
to a solution. When we run both multiMAGNA++ and GEDEVO-M for 100,000
generations, we observe the following (Figure
\ref{fig:time_mnc} and Figures \ref{fig:time_ciq} and \ref{fig:time_lccs}). In general,  GEDEVO-M converges slower than 
multiMAGNA++.  In fact, GEDEVO-M often does not converge even after
100,000 generations (Figure
\ref{fig:time_mnc}).  Thus, multiMAGNA++ can be stopped much
earlier than GEDEVO-M, while in the process mostly leading to superior
alignment quality.  On top of this, the time complexity of GEDEVO-M is
exponential with respect to the number of networks to be aligned as
opposed to the linear time complexity for multiMAGNA++ (Section
\ref{sec:complexity}).

Second, we compare running times of all methods. 
When we give the best case advantage to each method (i.e., when we run the parallelizable methods, multiMAGNA++ and GEDEVO-M, on multiple cores), multiMAGNA++ performs as follows (Figure \ref{fig:timing_ciq_best}). It is faster and more accurate than IsoRankN and FUSE on all data sets. Also, multiMAGNA++ is faster and more accurate than GEDEVO-M and BEAMS on at least one data set, while on the remaining data sets either it is comparable to these two methods both in terms of the running time and accuracy or it is slower but more accurate, in which case the increase in its accuracy justifies the increase in its running time. Regarding the remaining existing method, MI-Iso, multiMAGNA++ is slower but more accurate than MI-Iso on all data sets, which again justifies the increase in the running time compared to MI-Iso.
When we run all methods on a single core, as expected, both
multiMAGNA++ and GEDEVO-M slow down. Yet, for the largest network set
(PHY1), multiMAGNA++ is faster than IsoRankN and BEAMS while
typically being more accurate than these two methods and also than MI-Iso, GEDEVO-M, and FUSE
(Figures \ref{fig:timing_mnc}-\ref{fig:timing_lccs}). This is important, since
real-world networks will only continue to grow in size, and
multiMAGNA++ scales well to larger network data.


\section{Conclusion}
\label{sec:conclusion}

We present multiMAGNA++, an MNA extension of a state-of-the-art PNA
MAGNA++ that can directly optimize both node and edge conservation.
In general, multiMAGNA++ outperforms or is on par with the
existing MNA methods, most of which optimize node conservation only,
while often completing faster than the existing methods. 
That is, multiMAGNA++ scales well to larger network sizes and a larger number
of networks and can be parallelized effectively. In the
process of method evaluation, we introduce new alignment quality
measures for MNA to allow for more complete alignment characterization
as well as more fair MNA method evaluation compared to using only the
existing alignment quality measures, which may not fairly compare MNA
approaches that produce different output types (e.g., one-to-one
versus many-to-many node mappings). Thus, our study may impact future
MNA-related work in terms of both efficient method development and
fair method evaluation.


\section*{Acknowledgements}

We thank Lei Meng for providing the data used in this study.

\paragraph{Funding:}
This work was supported by National Science Foundation (NSF) CAREER CCF-1452795 and CCF-1319469 grants, and Air Force Young Investigator Program (AFOSR YIP).

\clearpage

\newcommand{\beginsupplement}{%
        \renewcommand{\thetable}{A\arabic{table}}%
        \renewcommand{\thefigure}{A\arabic{figure}}%
        \renewcommand{\thesection}{A\arabic{section}}%
}

\newcount\suppl
\suppl0
\def\Ref#1{\ifnum\suppl<1 \ref{#1} \else A\ref{#1}\fi}

\beginsupplement

\appendix

\section*{SUPPLEMENTARY SECTIONS}

\section{Introduction}
\label{sec:supplintro}

Examples of two-stage {\em PNA} methods are IsoRank \citep{IsoRank}, GHOST \citep{GHOST}, and the GRAAL family of methods \citep{GRAAL,CGRAAL,HGRAAL,MIGRAAL}.
IsoRank \citep{IsoRank} calculates node similarities using a PageRank-based spectral method and then uses a greedy alignment strategy.
GHOST \citep{GHOST} calculates node similarities by comparing ``spectral signatures'' of pairs of nodes.
GHOST then uses a two-phase alignment strategy consisting of a seed-and-extend global alignment stage followed by a local search procedure.
MI-GRAAL \citep{MIGRAAL}, the most recent and thus superior of all GRAAL family members, calculates node similarities using topological measures such as graphlet degree vector similarities (GDV-similarities) \citep{Milenkovic2008} and then maps nodes using a seed-and-extend alignment strategy.

Examples of two-stage {\em MNA} methods are IsoRankN \citep{IsoRankN}, MI-Iso \citep{Faisal2014}, SMETANA \citep{SMETANA}, BEAMS \citep{BEAMS}, NetCoffee \citep{NetCoffee}, CSRW \citep{RandomWalk} and FUSE \citep{FUSE}.
IsoRankN is among the first MNA methods to appear in the literature.
It calculates node similarities between all pairs of networks using IsoRank's node cost function and then creates an alignment by partitioning the graph of node similarities.
Recently, IsoRankN's node cost function was replaced with that of MI-GRAAL, thus resulting in a new method called MI-Iso \citep{Faisal2014}, which improved upon the original IsoRankN.
SMETANA calculates node similarities using a probabilistic model and then uses a greedy approach to align the networks.
BEAMS creates a graph of node similarities using protein sequence scores and then extracts from this graph a set of disjoint cliques that maximizes an alignment quality measure, in order to create a one-to-one alignment.
BEAMS then finds a many-to-many alignment by merging the cliques using an iterative greedy algorithm that maximizes the same alignment quality measure.
NetCoffee creates a weighted bipartite graph for every pair of networks by comparing sequence scores and neighborhood topologies of protein pairs. After calculating a one-to-one matching for each of the bipartite graphs, it uses a simulated annealing approach to contruct an MNA.
CSRW calculates node similarities using a context-sensitive random walk-based probabilistic model and then uses a greedy approach to align the networks.
FUSE calculates node similarities between all pairs of networks simultaneously using non-negative matrix tri-factorization \citep{NMTF} and then uses an approximate maximum weight $k$-partite matching algorithm to find an alignment between the multiple networks.

\section{Methods}
\label{sec:supplmethods}

\subsection{MultiMAGNA++}
\label{sec:supplmagna}

\subsubsection{Our representation of an MNA}
\label{sec:supplrepresentation}

Recall from Section \ref{sec:representation} in the main paper that a PNA of $G_1$ to $G_2$ is a total injective mapping $f \colon V_1 \mapsto V_2$; that is, every element in $V_1$ is matched uniquely with an element in $V_2$.
If $m = n$, then $f$ is a bijective mapping.
We need this constraint of $m = n$ to be satisfied in order to be able to represent a PNA as a permutation.
While in real life it is typically the case that $m < n$, we can easily impose the $m = n$ constraint, without making any special assumptions, by simply adding ``dummy'' zero-degree nodes, $z_i$, to $V_1$, so that ${\bar V_1} = V_1 \cup \{z_{m+1},z_{m+2},\ldots,z_n\}$.
In this way, we can simply assume that $m = n$ without explicitly referring to ${\bar V_1}$.

\subsubsection{Fitness function}
\label{sec:supplfitness}

Recall from Section \ref{sec:fitness} that when constructing topology+sequence alignments, we let multiMAGNA++ optimize (among other measures) BLAST sequence similarity as captured by E-value \citep{BLAST}, a commonly used node cost function for protein similarity \citep{IsoRankN,BEAMS,FUSE}.
Since E-value is a distance (rather than similarity) score,
and since multiMAGNA++ uses node similarities whose values should ideally lie between 0 and 1,
we transform each E-value to $-\log{(\textrm{E-value})}$ and then divide by the maximum of the transformed E-values.

\subsubsection{Tying the GA together}
\label{sec:supplparams}

Here, we expand our discussion from Section \ref{sec:params} in the main paper.
Up to this point in the main paper, we have discussed the components of our novel GA-based multiMAGNA++ that are needed to optimize the proposed fitness function using a population of MNAs.
In addition, other parameters of the GA include: 1) how to generate the initial population; 2) which population size to use; 3) how to choose which individuals of the population to cross; and 4) how many generations to run the algorithm for.
MAGNA++ used an initial population of randomly generated PNAs.
MAGNA++ also used initial populations that included alignments from the existing PNA methods.
Like MAGNA++ did for PNAs, we use an initial population consisting of randomly generated MNAs.
Since we represent an MNA of $k$ networks using $k-1$ permutations, a randomly generated MNA consists of $k-1$ randomly generated permutations.
While it is possible for multiMAGNA++ to add alignments from the existing MNA methods to its initial population, we did not consider this analysis in this work.
We do expect that using alignments from the existing MNA methods would result in further improvements in multiMAGNA++'s alignment quality.
Since GAs always perform better with larger population sizes \citep{Genetic}, we set population size to 15,000 MNAs, as was done by MAGNA++.
While it is possible to use an even larger population, even at the current population size, we see improvements of multiMAGNA++ over the existing methods.
While increasing the population size would likely further lead to even superior results of our method, this would also unnecessarily increase the method's running time.
In order to select parent alignments to be crossed, we use the roulette wheel selection algorithm, which chooses parents from the population of alignments with probability in proportion to the alignments' fitness.
The parent alignments are crossed in order to generate child alignments for the next generation, while keeping in the next generation a fraction of the best alignments from the previous generation.
The fraction of alignments we keep from the previous generation is 0.5.
We let multiMAGNA++ run for up to 100,000 generations.
This allows us to study the corresponding trends to determine an appropriate ``cut-off'' for stopping the algorithm.
We stop the algorithm when the fitness of the fittest alignment has increased less than 0.0001 in the last 500 generations.
The fittest alignment from the last generation is reported as multiMAGNA++'s final alignment.

\clearpage

\section*{SUPPLEMENTARY FIGURES}

\begin{figure}[h!]
\centering
\includegraphics[width=0.85\linewidth]{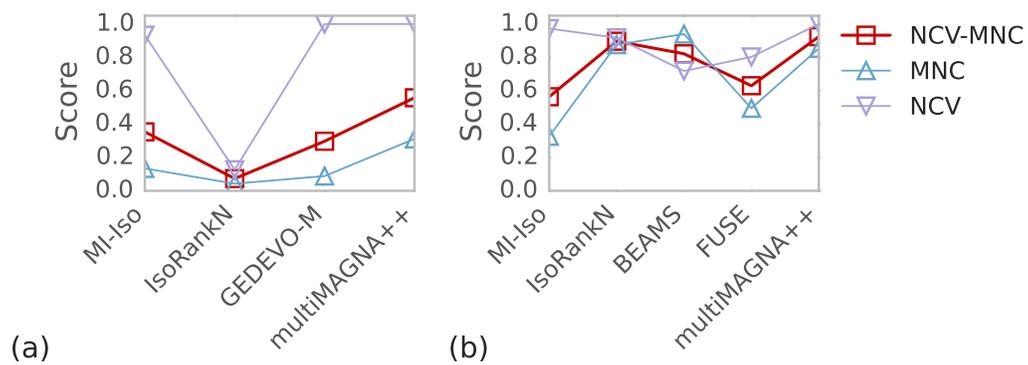}
\caption{NCV-MNC, NCV, and MNC for the Yeast+\%LC network set for \textbf{(a)} topology-only alignments and \textbf{(b)} topology+sequence alignments.}
\label{fig:mnc_yeastlc}
\end{figure}

%

\begin{figure}[h!]
\centering
\includegraphics[width=0.85\linewidth]{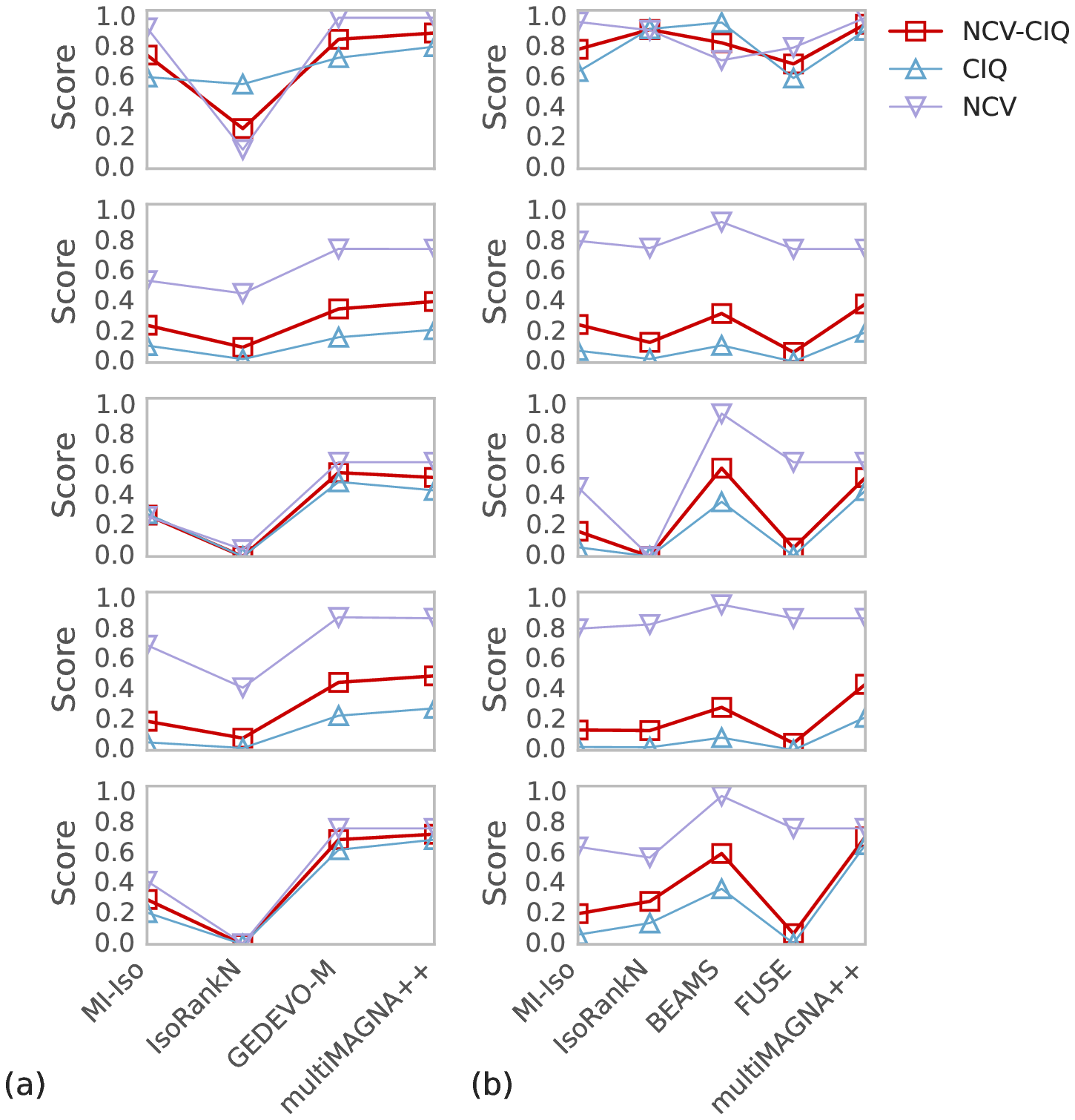}
\caption{NCV-CIQ, NCV, and CIQ for the five  network sets (Yeast+\%LC, PHY1, PHY2, Y2H1, and Y2H2 from top to bottom) for \textbf{(a)} topology-only alignments and \textbf{(b)} topology+sequence alignments.}
\label{fig:ciq}
\end{figure}

\begin{figure}[h!]
\centering
\includegraphics[width=0.85\linewidth]{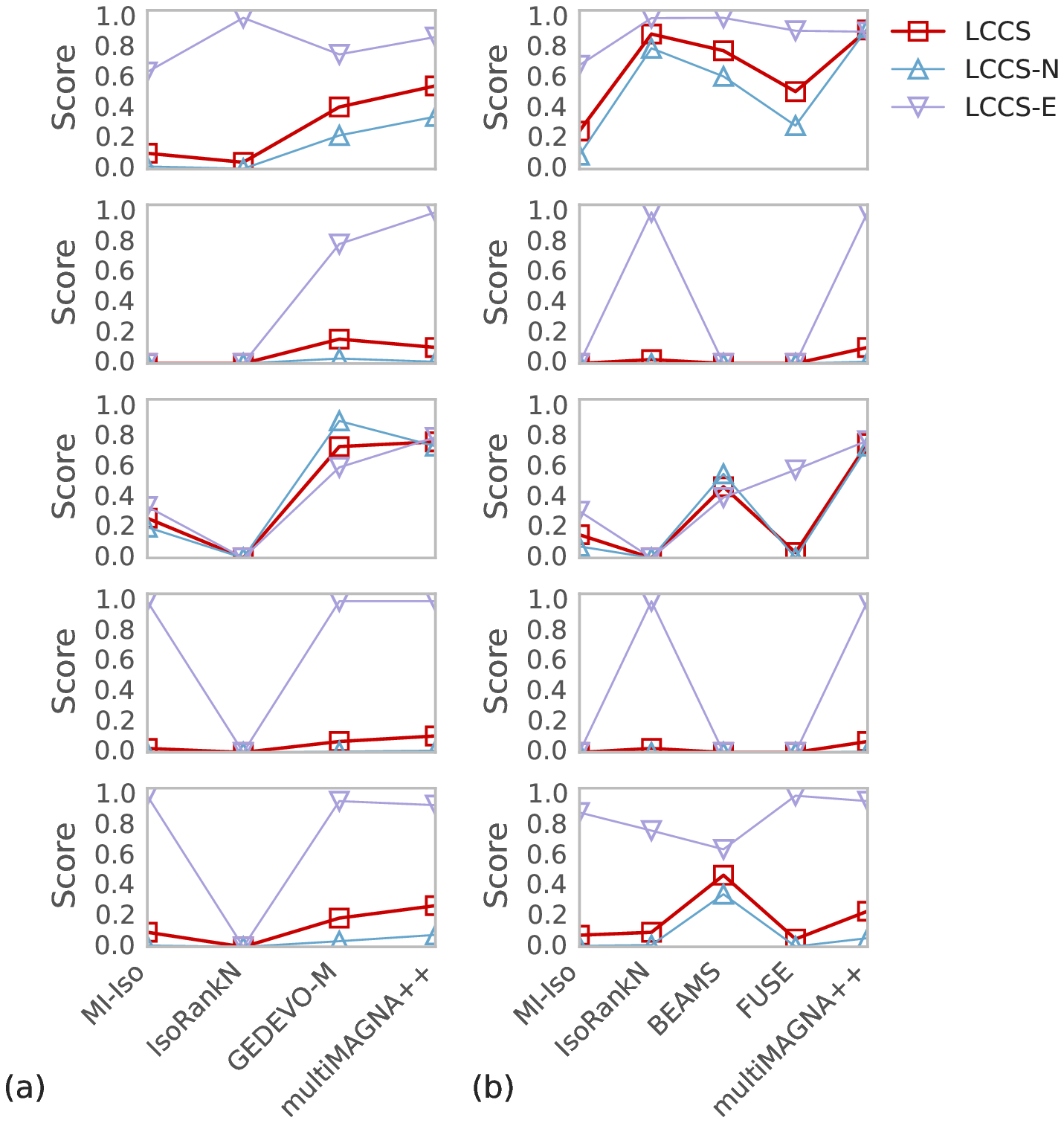}
\caption{LCCS for the five network sets (Yeast+\%LC, PHY1, PHY2, Y2H1, and Y2H2 from top to bottom) for \textbf{(a)} topology-only alignments and \textbf{(b)} topology+sequence alignments. }
\label{fig:lccs}
\end{figure}

\begin{figure}[h!]
\centering
\includegraphics[width=0.85\linewidth]{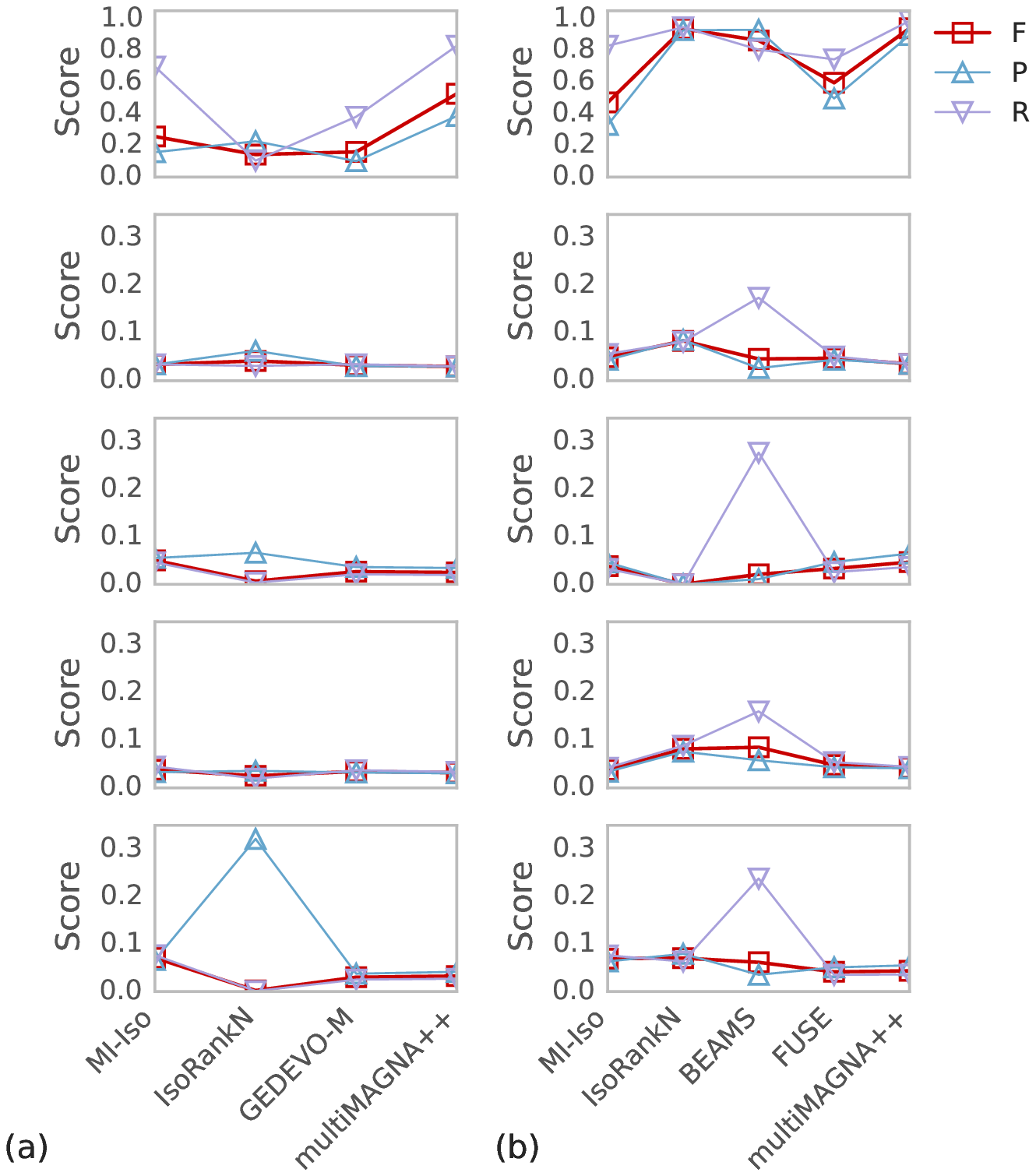}
\caption{F-score for the five network sets (Yeast+\%LC, PHY1, PHY2, Y2H1, and Y2H2 from top to bottom) for \textbf{(a)} topology-only alignments and \textbf{(b)} topology+sequence alignments.}
\label{fig:fscore}
\end{figure}

\begin{figure}[h!]
\centering
\includegraphics[width=0.73\linewidth]{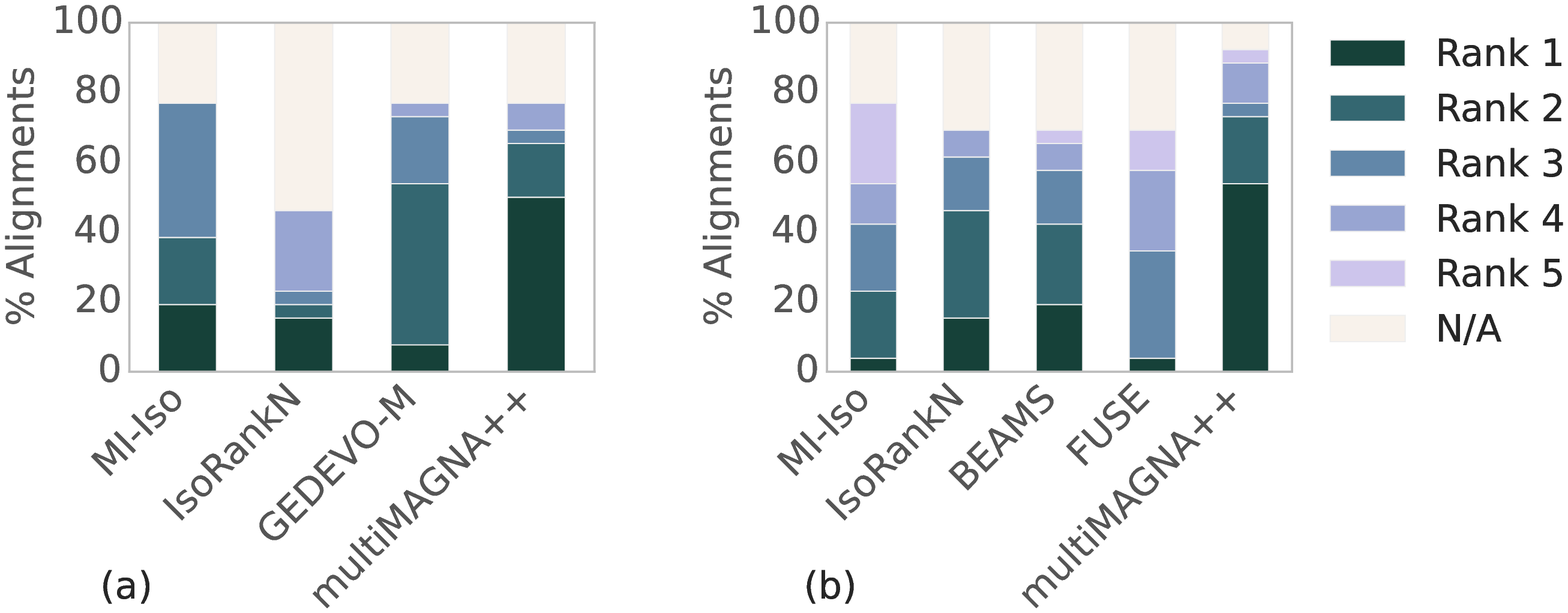}\\
\includegraphics[width=0.73\linewidth]{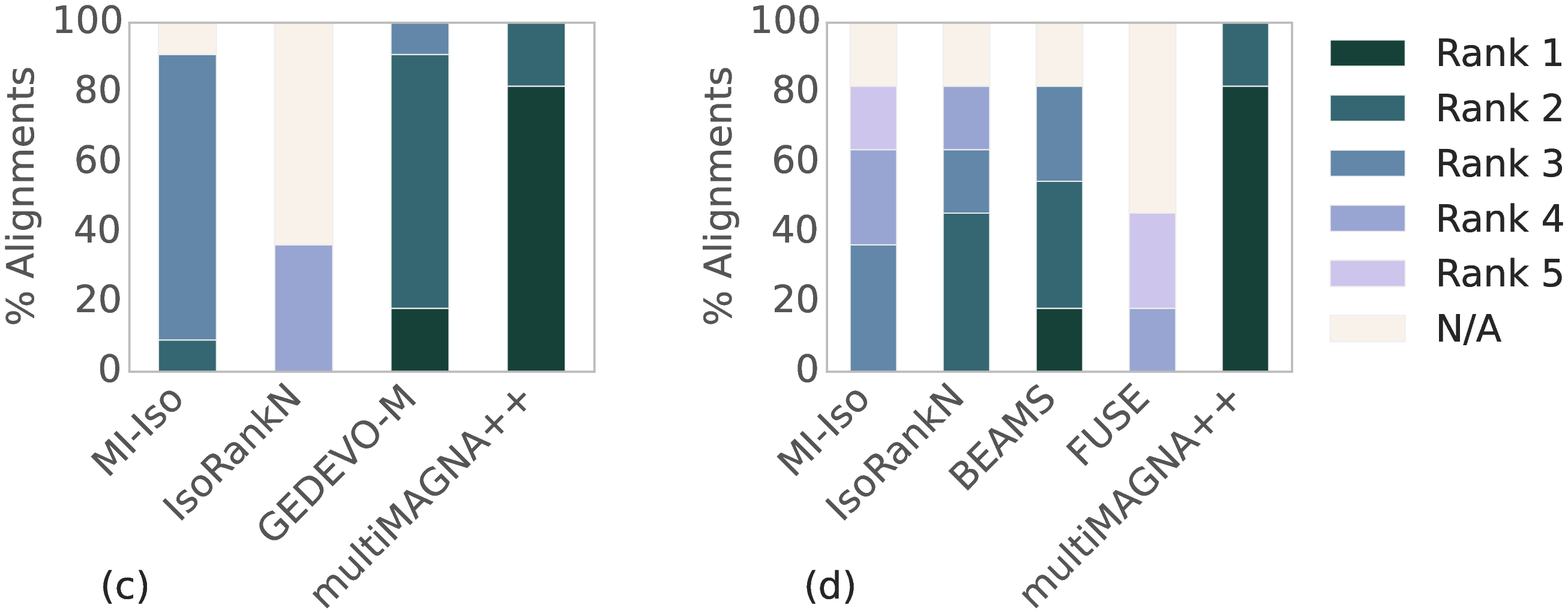}\\
\includegraphics[width=0.73\linewidth]{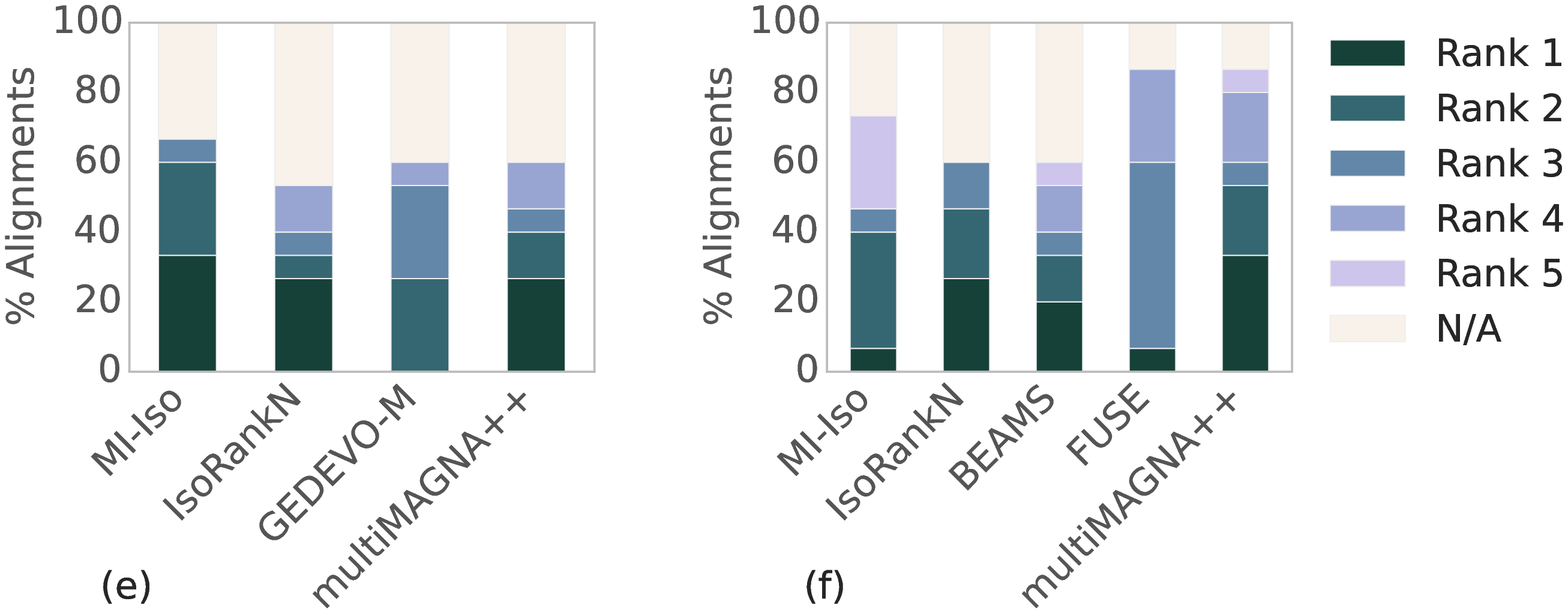}
\caption{Ranking of the MNA methods across all of Yeast+\%LC, PHY1, PHY2, Y2H1, and Y2H2 network sets with respect to \textbf{(a-b)} all measures,
\textbf{(c-d)} topological  NCV-MNC, NCV-CIQ, and
LCCS measures, and \textbf{(e-f)} functional MNE, GC, and F-score
measures, for \textbf{(a,c,e)} topology-only alignments and
\textbf{(b,d,f)} topology+sequence alignments. The ranking of each method 
is expressed as a percentage of all evaluation tests in which the
given method is the best performing (``Rank 1''), the second best
performing (``Rank 2''), etc. aligner of all considered methods. By
``best'', we mean the method with the lowest $p$-value, or the method with the best alignment quality score if the $p$-values are tied. If an alignment score of a method is not statistically
significant, the method is not ranked and is labelled as ``N/A''.}
\label{fig:all_rankings}
\end{figure}

\begin{figure}[h!]
\centering
\includegraphics[width=0.75\linewidth]{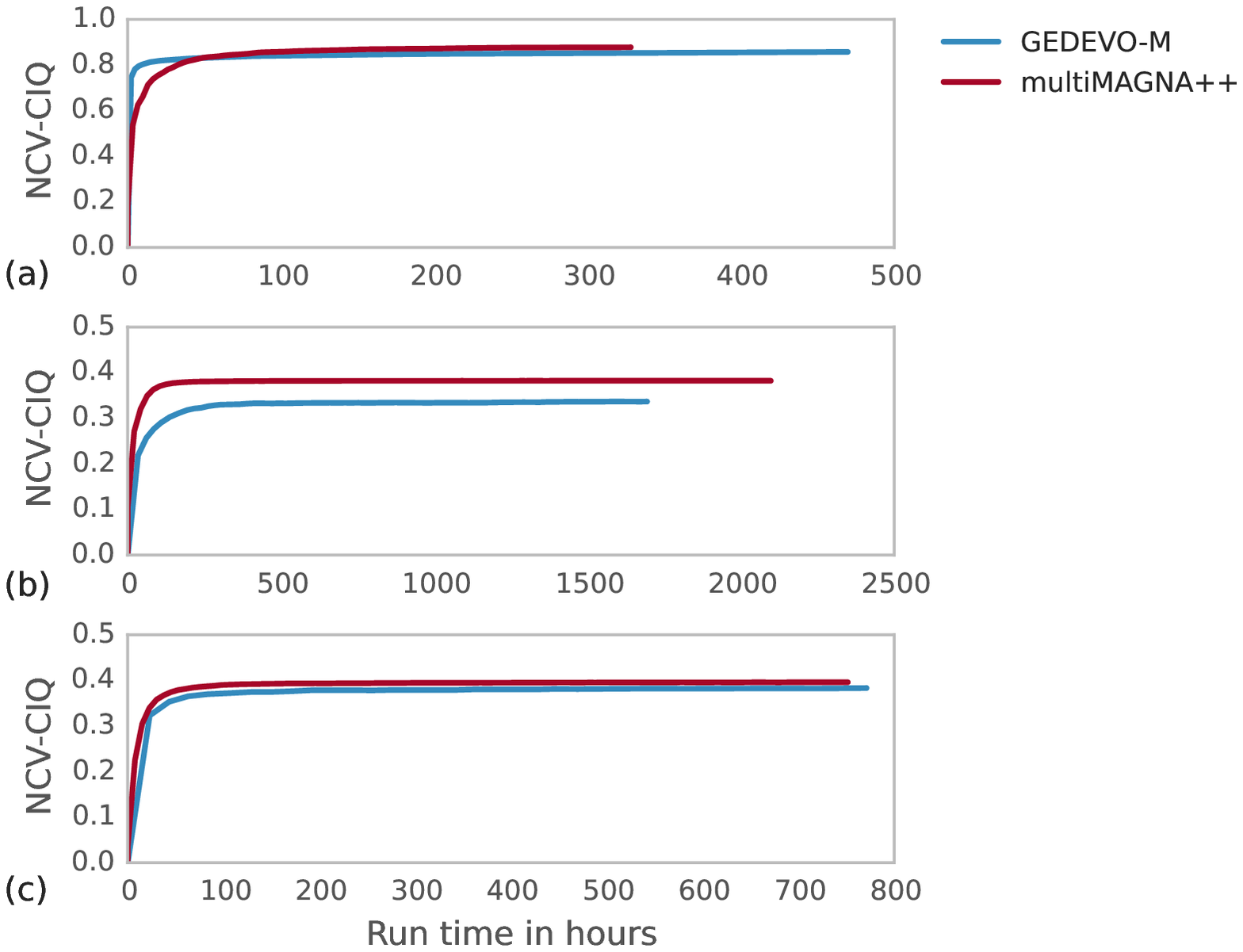}
\caption{NCV-CIQ as a function of the number of hours spent by multiMAGNA++ and GEDEVO-M when using a single thread, for topology-only alignments of \textbf{(a)} Yeast+\%LC, \textbf{(b)} PHY1, and \textbf{(c)} Y2H1  network sets. We only show these results for the three network sets with more than two networks; we leave out these results for PHY2 and Y2H2 that have two networks each.  Both multiMAGNA++ and GEDEVO-M  are run for 100,000 generations.}
\label{fig:time_ciq}
\end{figure}

\begin{figure}[h!]
\centering
\includegraphics[width=0.75\linewidth]{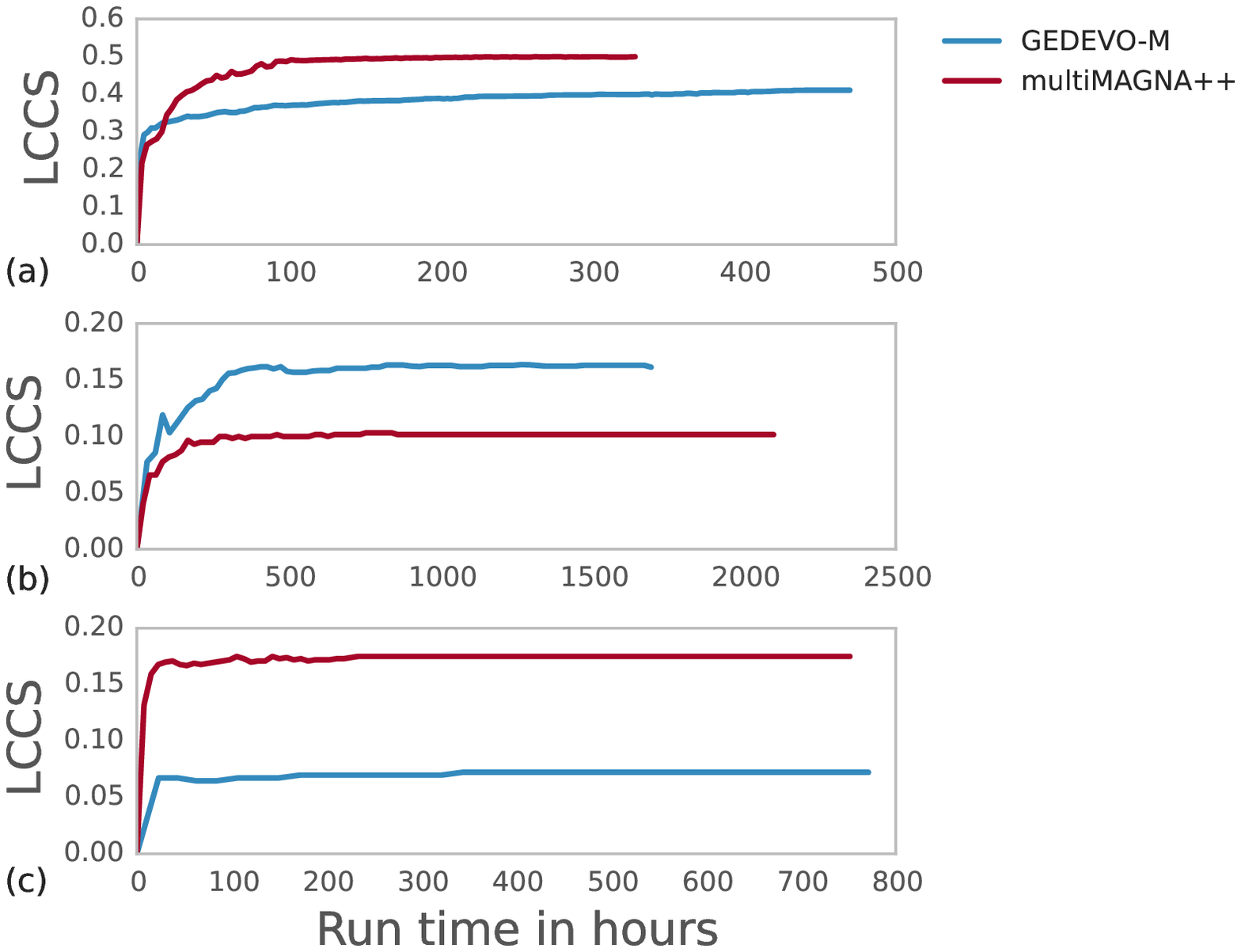}
\caption{LCCS as a function of the number of hours spent by multiMAGNA++ and GEDEVO-M when using a single thread, for topology-only alignments of \textbf{(a)} Yeast+\%LC, \textbf{(b)} PHY1, and \textbf{(c)} Y2H1  network sets. We only show these results for the three network sets with more than two networks; we leave out these results for PHY2 and Y2H2 that have two networks each. Both multiMAGNA++ and GEDEVO-M  are run for 100,000 generations. }
\label{fig:time_lccs}
\end{figure}

\begin{figure*}[h!]
\centering
\includegraphics[width=0.48\linewidth]{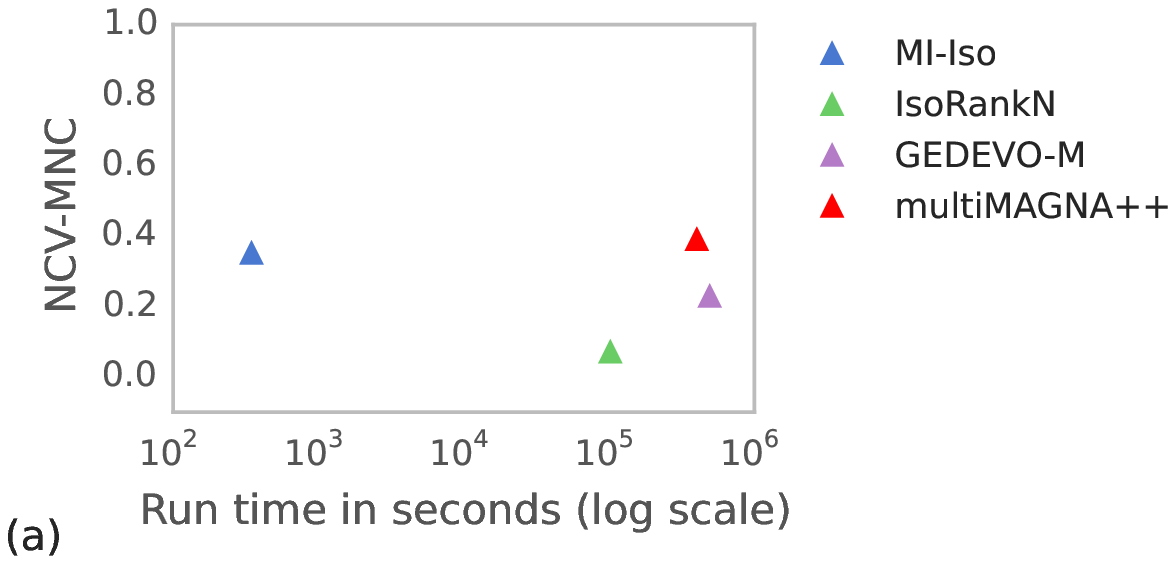}
\includegraphics[width=0.48\linewidth]{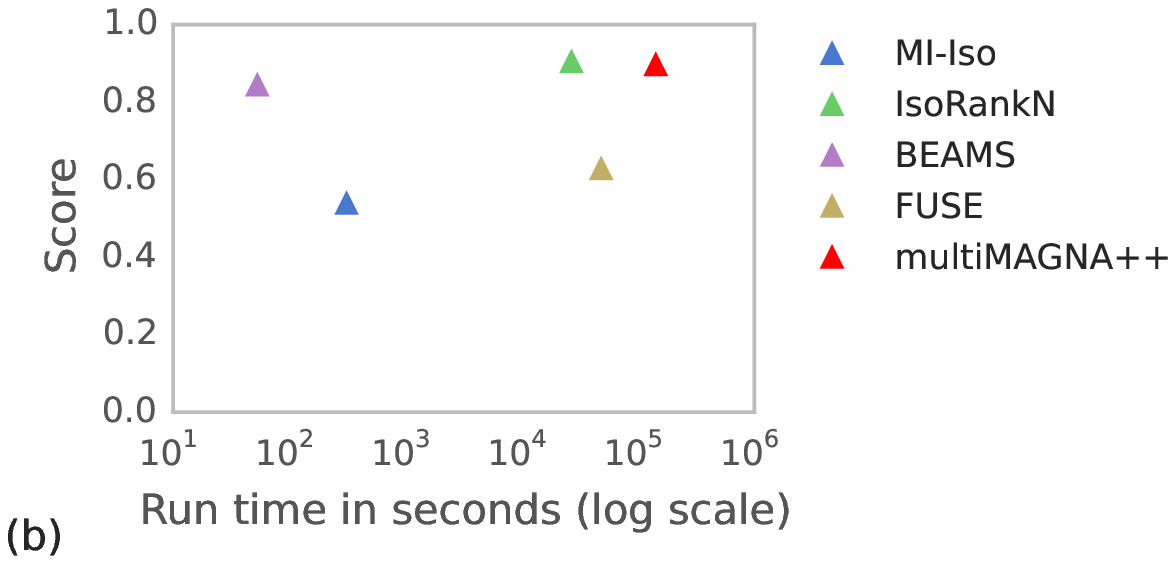}\\
\includegraphics[width=0.48\linewidth]{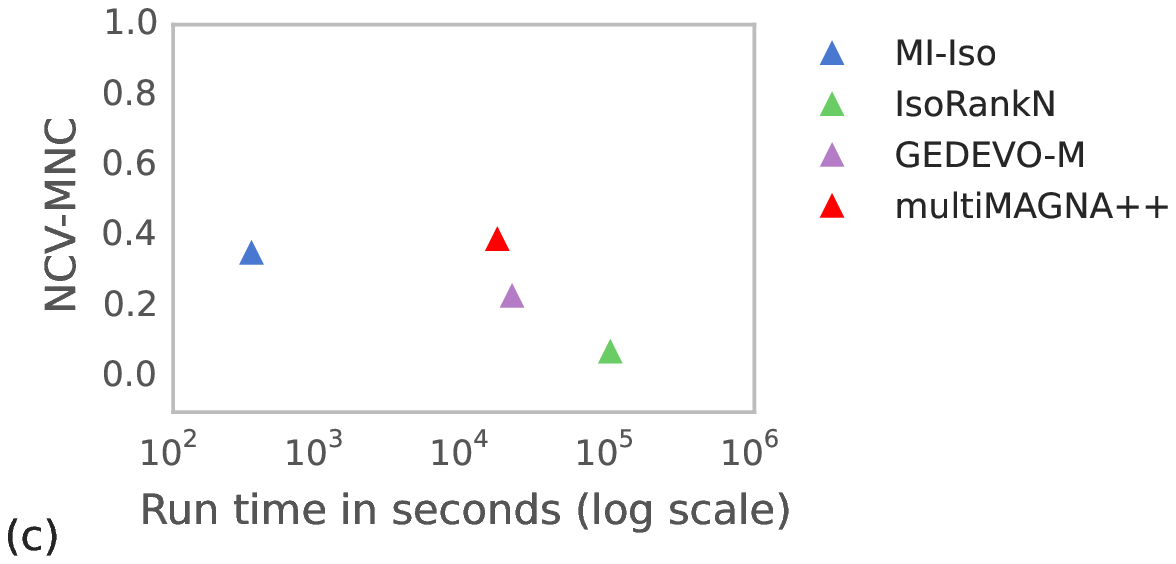}
\includegraphics[width=0.48\linewidth]{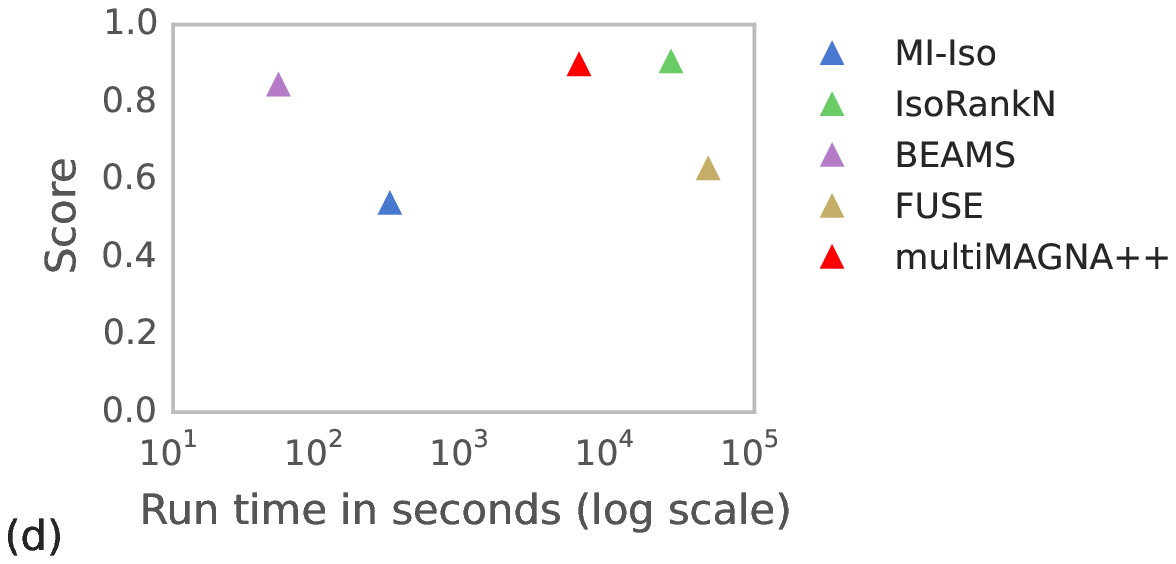}
\caption{NCV-MNC as a function of time when using \textbf{(a-b)}  a single thread and \textbf{(c-d)} 64 threads, for \textbf{(a,c)} topology-only alignments and \textbf{(b,d)} topology+sequence alignments, for the three network sets with more than two networks (we leave out these results for PHY2 and Y2H2 that have two networks each).}
\label{fig:timing_mnc}
\end{figure*}

\begin{figure*}[h!]
\centering
\includegraphics[width=0.48\linewidth]{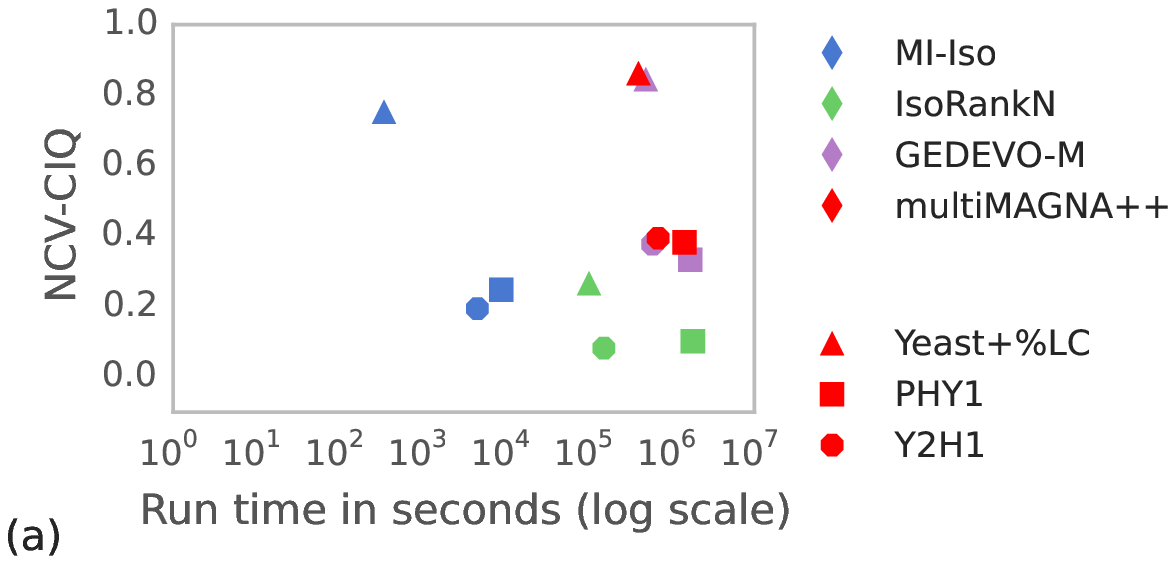}
\includegraphics[width=0.48\linewidth]{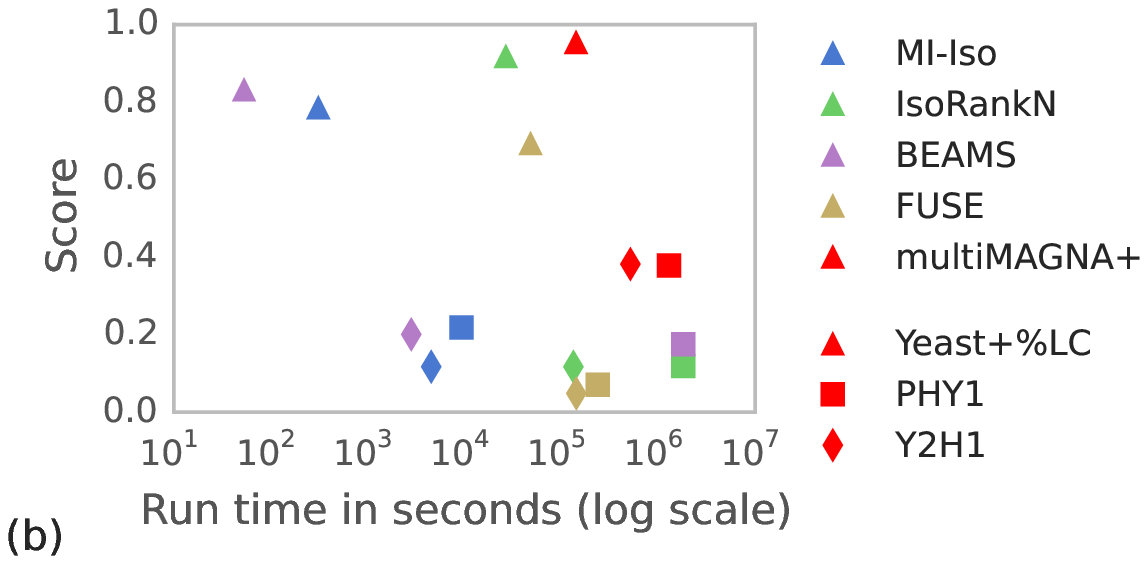}\\
\includegraphics[width=0.48\linewidth]{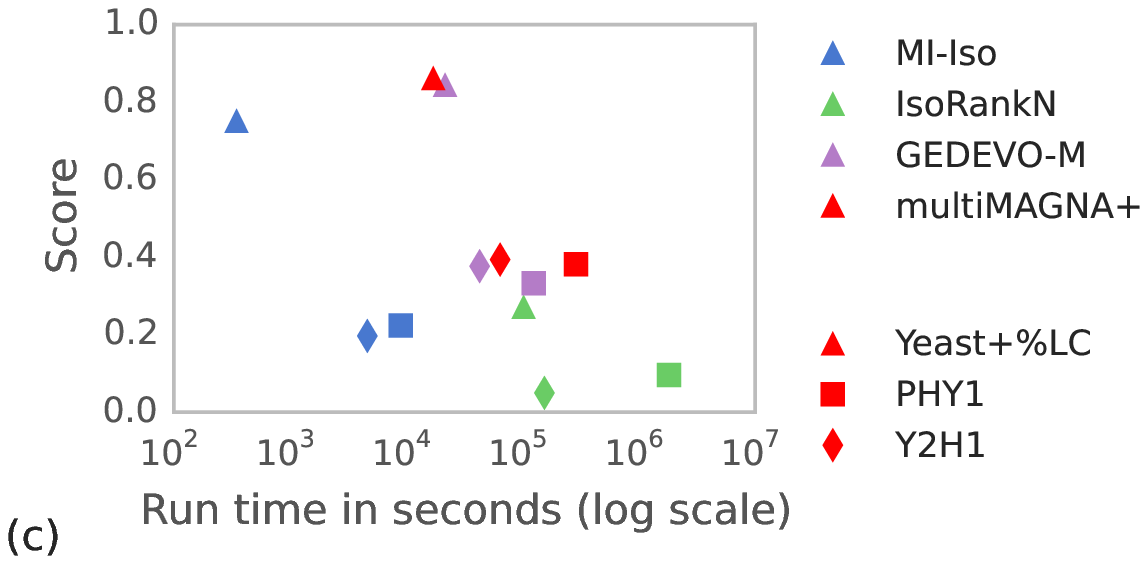}
\includegraphics[width=0.48\linewidth]{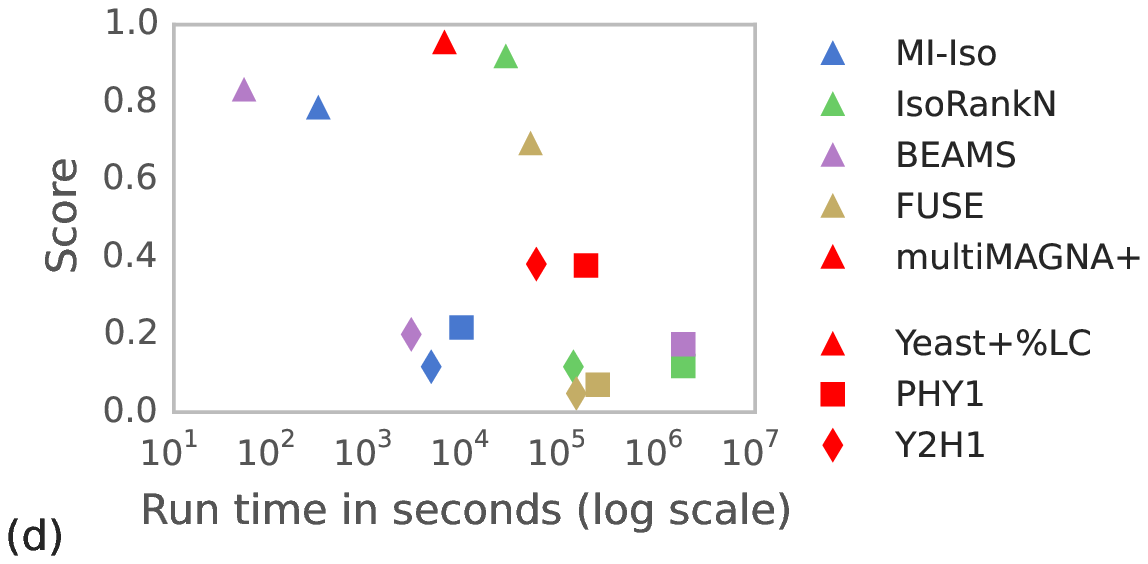}
\vspace{-0.4cm}
\caption{NCV-CIQ as a function of time when using \textbf{(a-b)} a single thread  and \textbf{(c-d)} 64 threads, for \textbf{(a,c)} topology-only alignments and \textbf{(b,d)} topology+sequence alignments, for the three network sets with more than two networks (we leave out these results for PHY2 and Y2H2 that have two networks each; Section \ref{sec:dataset}). For equivalent results for the remaining measures, see Figures S8 and S9.}
\label{fig:timing_ciq}
\end{figure*}

\begin{figure*}[h!]
\centering
\includegraphics[width=0.48\linewidth]{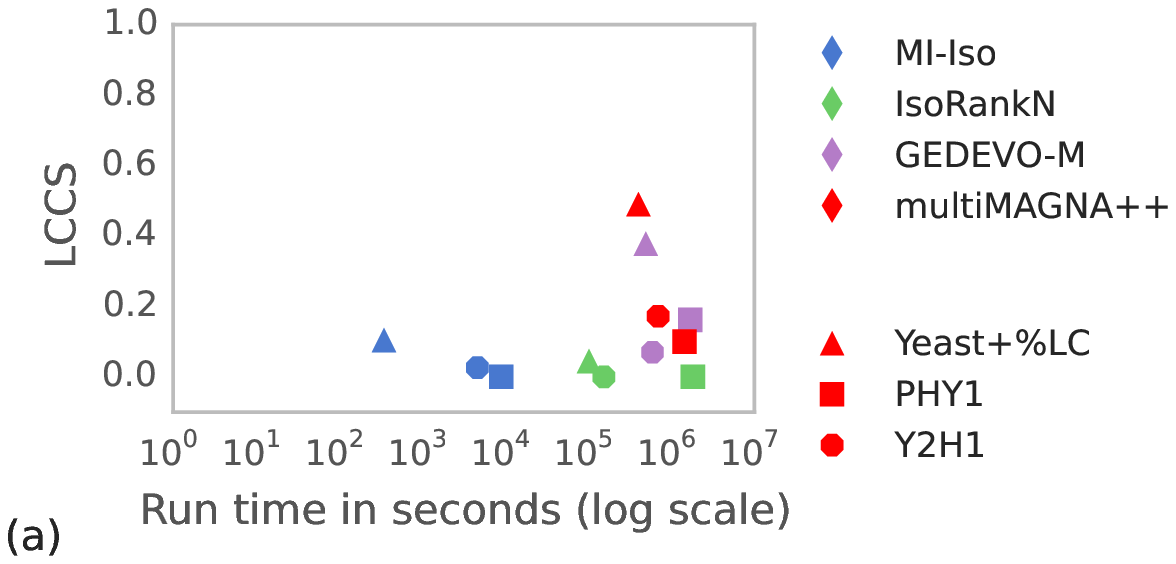}
\includegraphics[width=0.48\linewidth]{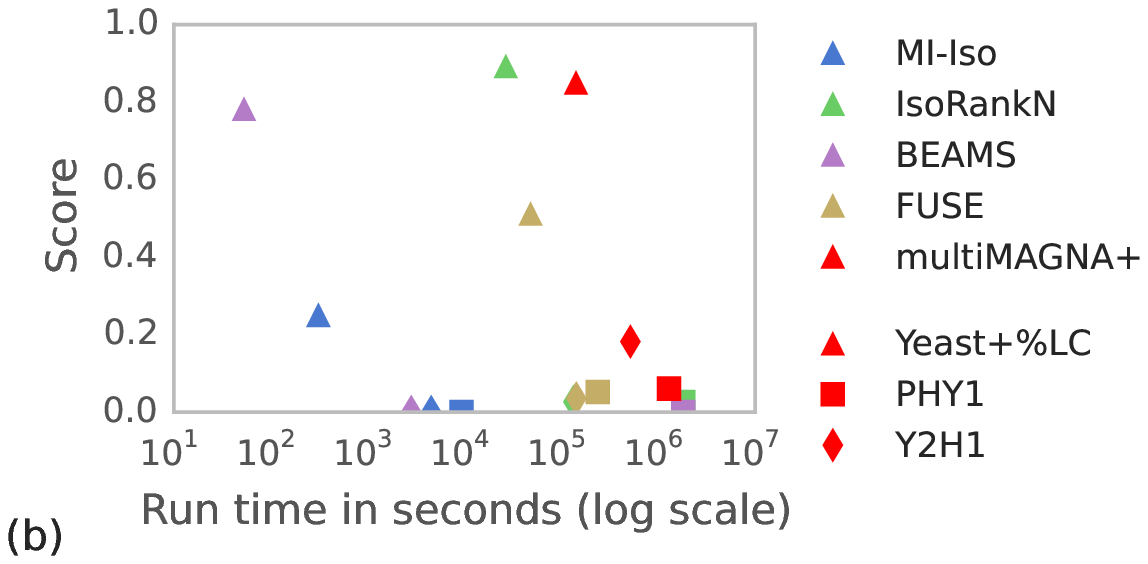}\\
\includegraphics[width=0.48\linewidth]{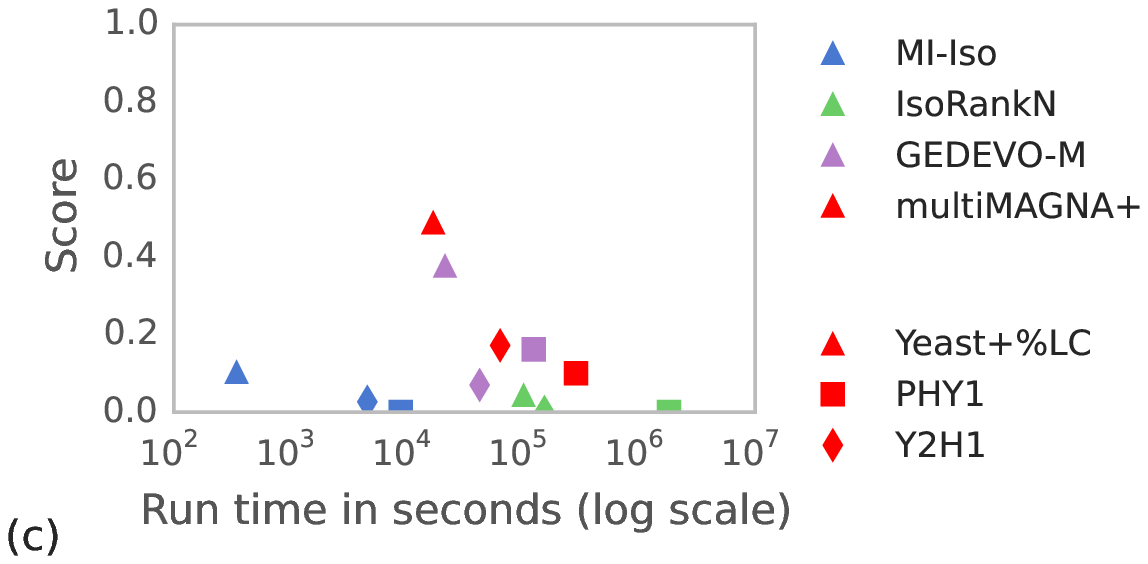}
\includegraphics[width=0.48\linewidth]{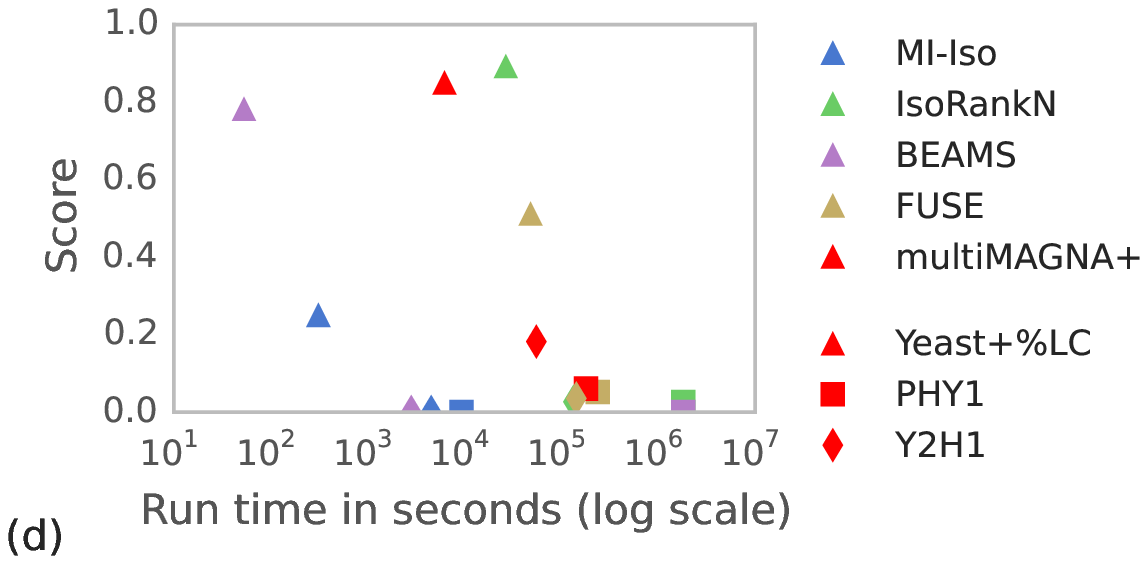}
\caption{LCCS as a function of time when using \textbf{(a-b)}  a single thread and \textbf{(c-d)} 64 threads, for \textbf{(a,c)} topology-only alignments and \textbf{(b,d)} topology+sequence alignments, for the three network sets with more than two networks (we leave out these results for PHY2 and Y2H2 that have two networks each). }
\label{fig:timing_lccs}
\end{figure*}

\clearpage
\section*{SUPPLEMENTARY TABLES}

\begin{table}[h!]
\begin{center}
\scriptsize
\begin{tabular}{|l|l|r|r|}
\hline
Set & Species & Proteins & Interactions \\ \hline
\multirow{6}{*}{Yeast+\%LC}
&                      Yeast+0\%LC   & 1,004 &  8,323 \\ \cline{2-4}
&                      Yeast+5\%LC   & 1,004 &  8,739 \\ \cline{2-4}
&                      Yeast+10\%LC  & 1,004 &  9,155 \\ \cline{2-4}
&                      Yeast+15\%LC  & 1,004 &  9,571 \\ \cline{2-4}
&                      Yeast+20\%LC  & 1,004 &  9,987 \\ \cline{2-4}
&                      Yeast+25\%LC  & 1,004 & 10,403 \\ \hline
\multirow{4}{*}{PHY1}
&                       Fly     & 7,887 & 36,285     \\ \cline{2-4}
&                       Worm    & 3,006 & 5,506      \\ \cline{2-4}
&                       Yeast   & 6,168 & 82,368     \\ \cline{2-4}
&                       Human   & 16,061 & 157,650   \\ \hline
\multirow{2}{*}{PHY2}
&                       Yeast   & 768 & 13,654     \\ \cline{2-4}
&                       Human   & 8,283 & 19,697   \\ \hline
\multirow{4}{*}{Y2H1}
&                       Fly     & 7,097 & 23,370   \\ \cline{2-4}
&                       Worm    & 2,874 & 5,199    \\ \cline{2-4}
&                       Yeast   & 3,427 & 11,348   \\ \cline{2-4}
&                       Human   & 9,996 & 39,984   \\ \hline
\multirow{2}{*}{Y2H2}
&                       Yeast   & 744 & 966       \\ \cline{2-4}
&                       Human   & 1,191 & 1,567   \\ \hline
\end{tabular}
\normalsize
\end{center}
\caption{The five PPI networks sets that we use in our study. We know the true mapping for the Yeast+\%LC network set unlike for the other network sets. For PHY2 and Y2H2, we only use the yeast and human PPI networks since the fly and worm networks are too small for analysis.}
\label{tab:nets}
\end{table}

\begin{table}[h!]
\begin{center}
\scriptsize
\begin{tabular}{|l|l|}
\hline
Algorithms & Parameters  \\ \hline
\multicolumn{2}{|l|}{Topology-only alignments} \\ \hline
IsoRankN       & K=30 thresh=1$\text{\sc{e}-}$4 maxveclen=5000000 alpha=1.0 \\ \hline
MI-Iso         & K=30 thresh=1$\text{\sc{e}-}$4 maxveclen=5000000 alpha=1.0  \\ \hline
GEDEVO-M       & beta=0.4 \\ \hline
multiMAGNA++    & m=CIQ p=15000 n=100000 e=0.5 a=1.0   \\ \hline
\multicolumn{2}{|l|}{Topology+sequence alignments} \\ \hline
IsoRankN      & K=30 thresh=1$\text{\sc{e}-}$4 maxveclen=5000000 alpha=0.5   \\ \hline
MI-Iso        & K=30 thresh=1$\text{\sc{e}-}$4 maxveclen=5000000 alpha=1.0  \\ \hline
BEAMS         & beta=0.4 alpha=0.5 \\ \hline
FUSE          & k=[100,100,100,100] iter\_num=1000 gamma=0.7 a=0.5 \\ \hline
multiMAGNA++  & m=CIQ p=15000 n=100000 e=0.5 a=0.25 \\ \hline
\end{tabular}
\normalsize
\end{center}
\caption{Parameters of the MNA methods. For the existing methods, we use parameters that were recommended in the methods' original publications. The parameters of multiMAGNA++ are the edge conservation measure (m), population size (p), number of generations (n), and the $\alpha$ parameter (a).}
\label{tab:params}
\end{table}

\begin{table}[h!]
\begin{center}
\scriptsize
\begin{tabular}{|p{22mm}|p{18mm}|p{18mm}|p{18mm}|p{18mm}|p{18mm}|p{18mm}|} \hline
 & \multicolumn{3}{c|}{Topological measures} & \multicolumn{3}{c|}{Functional measures} \\ \hline
   Method
 & NCV-MNC
 & NCV-CIQ
 & LCCS
 & MNE
 & GC
 & F-score
 \\ \hline
 MI-Iso
 & 0.3533 \newline p $<$ 1.00$\text{e-}$4
 & 0.7526 \newline p $<$ 1.00$\text{e-}$4
 & 0.1042 \newline p $<$ 1.00$\text{e-}$4
 & 0.9578 \newline p $<$ 1.00$\text{e-}$4
 & 0.5143 \newline p $<$ 1.00$\text{e-}$4
 & 0.2558 \newline p $<$ 1.00$\text{e-}$4
 \\ \cline{1-7}
 IsoRankN
 & 0.0730 \newline p $<$ 1.00$\text{e-}$4
 & 0.2655 \newline p $<$ 1.00$\text{e-}$4
 & {\it \textcolor{gray!125} { 0.0446 \newline p = 0.026 }}
 & 0.9630 \newline p $<$ 1.00$\text{e-}$4
 & 0.7376 \newline p $<$ 1.00$\text{e-}$4
 & 0.1420 \newline p $<$ 1.00$\text{e-}$4
 \\ \cline{1-7}
 GEDEVO-M
 & 0.2969 \newline p $<$ 1.00$\text{e-}$4
 & 0.8574 \newline p $<$ 1.00$\text{e-}$4
 & 0.4104 \newline p $<$ 1.00$\text{e-}$4
 & 0.9610 \newline p $<$ 1.00$\text{e-}$4
 & 0.4333 \newline p $<$ 1.00$\text{e-}$4
 & 0.1586 \newline p $<$ 1.00$\text{e-}$4
 \\ \cline{1-7}
 multiMAGNA++
 & {\bf 0.5578 \newline p $<$ 1.00$\textbf{e-}$4 }
 & {\bf 0.8991 \newline p $<$ 1.00$\textbf{e-}$4 }
 & {\bf 0.5501 \newline p $<$ 1.00$\textbf{e-}$4 }
 & {\bf 0.9341 \newline p $<$ 1.00$\textbf{e-}$4 }
 & {\bf 0.7739 \newline p $<$ 1.00$\textbf{e-}$4 }
 & {\bf 0.5258 \newline p $<$ 1.00$\textbf{e-}$4 }
 \\ \hline
\end{tabular}
\normalsize
\end{center}
\caption{Alignment accuracy of different MNA approaches for the Yeast+\%LC network set in terms of  topological NCV-MNC, NCV-CIQ, and LCCS measures and functional MNE, GO correctness (GC), and F-score measures, for topology-only alignments. The symbol ``p'' signifies $p$-values of the observed alignment scores, as defined in Section \ref{sec:othermethods}. For each alignment quality measure (i.e., in each column), the best method (i.e., the method with the lowest $p$-value, or the method with the best alignment quality score if the $p$-values are tied) is bolded. The alignment scores that are not statistically significant, if any, are greyed out and italicized. Note that for MNE, the lower the score, the better the alignment quality. For all other measures, the higher the score, the better the alignment quality.} 
\label{tab:all_yeastlc_top}
\end{table}

\begin{table}[h!]
\begin{center}
\scriptsize
\begin{tabular}{|p{22mm}|p{18mm}|p{18mm}|p{18mm}|p{18mm}|p{18mm}|p{18mm}|} \hline
 & \multicolumn{3}{c|}{Topological measures} & \multicolumn{3}{c|}{Functional measures} \\ \hline
   Method
 & NCV-MNC
 & NCV-CIQ
 & LCCS
 & MNE
 & GC
 & F-score
 \\ \hline
 MI-Iso
 & 0.5638 \newline p $<$ 1.00$\text{e-}$4
 & 0.7908 \newline p $<$ 1.00$\text{e-}$4
 & 0.2508 \newline p $<$ 1.00$\text{e-}$4
 & 0.9508 \newline p $<$ 1.00$\text{e-}$4
 & 0.6137 \newline p $<$ 1.00$\text{e-}$4
 & 0.4709 \newline p $<$ 1.00$\text{e-}$4
 \\ \cline{1-7}
 IsoRankN
 & 0.8967 \newline p $<$ 1.00$\text{e-}$4
 & 0.9219 \newline p $<$ 1.00$\text{e-}$4
 & 0.8929 \newline p $<$ 1.00$\text{e-}$4
 & 0.9532 \newline p $<$ 1.00$\text{e-}$4
 & 0.9748 \newline p $<$ 1.00$\text{e-}$4
 & 0.9367 \newline p $<$ 1.00$\text{e-}$4
 \\ \cline{1-7}
 BEAMS
 & 0.8215 \newline p $<$ 1.00$\text{e-}$4
 & 0.8342 \newline p $<$ 1.00$\text{e-}$4
 & 0.7827 \newline p $<$ 1.00$\text{e-}$4
 & 0.9667 \newline p $<$ 1.00$\text{e-}$4
 & 0.9882 \newline p $<$ 1.00$\text{e-}$4
 & 0.8625 \newline p $<$ 1.00$\text{e-}$4
 \\ \cline{1-7}
 FUSE
 & 0.6299 \newline p $<$ 1.00$\text{e-}$4
 & 0.6943 \newline p $<$ 1.00$\text{e-}$4
 & 0.5123 \newline p $<$ 1.00$\text{e-}$4
 & 0.9618 \newline p $<$ 1.00$\text{e-}$4
 & 0.7744 \newline p $<$ 1.00$\text{e-}$4
 & 0.5953 \newline p $<$ 1.00$\text{e-}$4
 \\ \cline{1-7}
 multiMAGNA++
 & {\bf 0.9241 \newline p $<$ 1.00$\textbf{e-}$4 }
 & {\bf 0.9574 \newline p $<$ 1.00$\textbf{e-}$4 }
 & {\bf 0.9201 \newline p $<$ 1.00$\textbf{e-}$4 }
 & {\bf 0.9347 \newline p $<$ 1.00$\textbf{e-}$4 }
 & {\bf 0.9897 \newline p $<$ 1.00$\textbf{e-}$4 }
 & {\bf 0.9392 \newline p $<$ 1.00$\textbf{e-}$4 }
 \\ \hline
\end{tabular}
\normalsize
\end{center}
\caption{Alignment accuracy of different MNA approaches for the Yeast+\%LC network set in terms of topological NCV-MNC, NCV-CIQ, and LCCS measures and functional MNE, GO correctness (GC), and F-score measures, for topology+sequence alignments. The symbol ``p'' signifies $p$-values of the observed alignment scores. For each alignment quality measure (i.e., in each column), the best method (i.e., the method with the lowest $p$-value, or the method with the best alignment quality score if the $p$-values are tied) is bolded. The alignment scores that are not statistically significant, if any, are greyed out and italicized. Note that for MNE, the lower the score, the better the alignment quality. For all other measures, the higher the score, the better the alignment quality.}
\label{tab:all_yeastlc_topseq}
\end{table}

\begin{table}[h!]
\begin{center}
\scriptsize
\begin{tabular}{|l|p{22mm}|p{18mm}|p{18mm}|p{18mm}|p{18mm}|p{18mm}|} \hline
 & & \multicolumn{2}{c|}{Topological measures} & \multicolumn{3}{c|}{Functional measures} \\ \hline
 & Method
 & NCV-CIQ
 & LCCS
 & MNE
 & GC
 & F-score
 \\ \hline
\parbox[t]{2mm}{\multirow{12}{*}{\rotatebox[origin=c]{90}{PHY2}}}
 & MI-Iso
 & 0.2746 \newline p $<$ 1.00$\text{e-}$4
 & 0.2632 \newline p $<$ 1.00$\text{e-}$4
 & 0.9736 \newline p = 3.00\text{e-}4
 & {\bf 0.4034 \newline p $<$ 1.00$\textbf{e-}$4 }
 & {\bf 0.0503 \newline p $<$ 1.00$\textbf{e-}$4 }
 \\ \cline{2-7}
 & IsoRankN
 & {\it \textcolor{gray!125} { 0.0000 \newline p = 1.000 }}
 & {\it \textcolor{gray!125} { 0.0000 \newline p = 1.000 }}
 & {\it \textcolor{gray!125} { 1.0000 \newline p = 0.376 }}
 & {\it \textcolor{gray!125} { 0.2521 \newline p = 0.009 }}
 & {\it \textcolor{gray!125} { 0.0067 \newline p = 0.382 }}
 \\ \cline{2-7}
 & BEAMS
 & {\bf 0.5858 \newline p $<$ 1.00$\textbf{e-}$4 }
 & 0.4714 \newline p $<$ 1.00$\text{e-}$4
 & {\bf 0.9799 \newline p $<$ 1.00$\textbf{e-}$4 }
 & {\it \textcolor{gray!125} { 0.5833 \newline p = 1.000 }}
 & 0.0210 \newline p $<$ 1.00$\text{e-}$4
 \\ \cline{2-7}
 & FUSE
 & {\it \textcolor{gray!125} { 0.0561 \newline p = 0.002 }}
 & {\it \textcolor{gray!125} { 0.0352 \newline p = 0.249 }}
 & {\it \textcolor{gray!125} { 0.9808 \newline p = 0.271 }}
 & 0.2422 \newline p $<$ 1.00$\text{e-}$4
 & 0.0332 \newline p $<$ 1.00$\text{e-}$4
 \\ \cline{2-7}
 & GEDEVO-M
 & 0.5562 \newline p $<$ 1.00$\text{e-}$4
 & 0.7369 \newline p $<$ 1.00$\text{e-}$4
 & {\it \textcolor{gray!125} { 0.9786 \newline p = 0.042 }}
 & 0.2096 \newline p $<$ 1.00$\text{e-}$4
 & 0.0266 \newline p $<$ 1.00$\text{e-}$4
 \\ \cline{2-7}
 & multiMAGNA++
 & 0.5236 \newline p $<$ 1.00$\text{e-}$4
 & {\bf 0.7680 \newline p $<$ 1.00$\textbf{e-}$4 }
 & 0.9745 \newline p = 1.00\text{e-}4
 & 0.3033 \newline p $<$ 1.00$\text{e-}$4
 & 0.0466 \newline p $<$ 1.00$\text{e-}$4
 \\ \hline
\parbox[t]{2mm}{\multirow{12}{*}{\rotatebox[origin=c]{90}{Y2H2}}}
 & MI-Iso
 & 0.2972 \newline p $<$ 1.00$\text{e-}$4
 & 0.0970 \newline p $<$ 1.00$\text{e-}$4
 & {\it \textcolor{gray!125} { 0.9880 \newline p = 0.313 }}
 & {\it \textcolor{gray!125} { 0.5567 \newline p = 0.131 }}
 & {\bf 0.0710 \newline p $<$ 1.00$\textbf{e-}$4 }
 \\ \cline{2-7}
 & IsoRankN
 & 0.2853 \newline p $<$ 1.00$\text{e-}$4
 & 0.0965 \newline p $<$ 1.00$\text{e-}$4
 & {\it \textcolor{gray!125} { 0.9885 \newline p = 0.641 }}
 & {\bf 0.3651 \newline p $<$ 1.00$\textbf{e-}$4 }
 & 0.0706 \newline p $<$ 1.00$\text{e-}$4
 \\ \cline{2-7}
 & BEAMS
 & 0.6020 \newline p $<$ 1.00$\text{e-}$4
 & {\bf 0.4741 \newline p $<$ 1.00$\textbf{e-}$4 }
 & {\it \textcolor{gray!125} { 0.9869 \newline p = 0.057 }}
 & {\it \textcolor{gray!125} { 0.8222 \newline p = 0.783 }}
 & 0.0616 \newline p $<$ 1.00$\text{e-}$4
 \\ \cline{2-7}
 & FUSE
 & {\it \textcolor{gray!125} { 0.0707 \newline p = 0.028 }}
 & {\it \textcolor{gray!125} { 0.0518 \newline p = 0.889 }}
 & {\it \textcolor{gray!125} { 0.9949 \newline p = 0.644 }}
 & 0.2801 \newline p $<$ 1.00$\text{e-}$4
 & 0.0416 \newline p $<$ 1.00$\text{e-}$4
 \\ \cline{2-7}
 & GEDEVO-M
 & 0.6945 \newline p $<$ 1.00$\text{e-}$4
 & 0.1905 \newline p $<$ 1.00$\text{e-}$4
 & {\it \textcolor{gray!125} { 0.9968 \newline p = 0.926 }}
 & {\it \textcolor{gray!125} { 0.2228 \newline p = 0.714 }}
 & {\it \textcolor{gray!125} { 0.0303 \newline p = 0.486 }}
 \\ \cline{2-7}
 & multiMAGNA++
 & {\bf 0.7307 \newline p $<$ 1.00$\textbf{e-}$4 }
 & 0.2727 \newline p $<$ 1.00$\text{e-}$4
 & {\it \textcolor{gray!125} { 0.9908 \newline p = 0.090 }}
 & 0.2920 \newline p $<$ 1.00$\text{e-}$4
 & 0.0436 \newline p $<$ 1.00$\text{e-}$4
 \\ \hline
\end{tabular}
\normalsize
\end{center}
\caption{Alignment accuracy selected of different MNA approaches for the PHY2 and Y2H2 network sets in terms of topological NCV-CIQ and LCCS measures and functional MNE, GO correctness (GC), and F-score measures. The symbol ``p'' signifies $p$-values of the observed alignment scores. For each alignment quality measure, and for each network set, the best method (i.e., the method with the lowest $p$-value, or the method with the best alignment quality score if the $p$-values are tied) is bolded. The alignment scores that are not statistically significant, if any, are greyed out and italicized. Note that for MNE, the lower the score, the better the alignment quality. For all other measures, the higher the score, the better the alignment quality.}
\label{tab:all_all_best}
\end{table}

\begin{table}[h!]
\begin{center}
\scriptsize
\begin{tabular}{|l|p{22mm}|p{18mm}|p{18mm}|p{18mm}|p{18mm}|p{18mm}|} \hline
 & & \multicolumn{2}{c|}{Topological measures} & \multicolumn{3}{c|}{Functional measures} \\ \hline
 & Method
 & NCV-CIQ
 & LCCS
 & MNE
 & GC
 & F-score
 \\ \hline
\parbox[t]{2mm}{\multirow{8}{*}{\rotatebox[origin=c]{90}{PHY1}}}
 & MI-Iso
 & 0.2472 \newline p $<$ 1.00$\text{e-}$4
 & {\it \textcolor{gray!125} { 0.0000 \newline p = 1.000 }}
 & {\it \textcolor{gray!125} { 0.8912 \newline p = 0.579 }}
 & 0.1913 \newline p $<$ 1.00$\text{e-}$4
 & 0.0341 \newline p $<$ 1.00$\text{e-}$4
 \\ \cline{2-7}
 & IsoRankN
 & 0.1012 \newline p $<$ 1.00$\text{e-}$4
 & {\it \textcolor{gray!125} { 0.0000 \newline p = 1.000 }}
 & {\bf 0.7977 \newline p $<$ 1.00$\textbf{e-}$4 }
 & {\bf 0.2535 \newline p $<$ 1.00$\textbf{e-}$4 }
 & {\bf 0.0415 \newline p $<$ 1.00$\textbf{e-}$4 }
 \\ \cline{2-7}
 & GEDEVO-M
 & 0.3554 \newline p $<$ 1.00$\text{e-}$4
 & {\bf 0.1613 \newline p $<$ 1.00$\textbf{e-}$4 }
 & {\it \textcolor{gray!125} { 0.9205 \newline p = 1.000 }}
 & 0.1721 \newline p $<$ 1.00$\text{e-}$4
 & 0.0324 \newline p $<$ 1.00$\text{e-}$4
 \\ \cline{2-7}
 & multiMAGNA++
 & {\bf 0.4046 \newline p $<$ 1.00$\textbf{e-}$4 }
 & 0.1064 \newline p $<$ 1.00$\text{e-}$4
 & 0.8449 \newline p $<$ 1.00$\text{e-}$4
 & 0.1610 \newline p $<$ 1.00$\text{e-}$4
 & 0.0299 \newline p $<$ 1.00$\text{e-}$4
 \\ \hline
\parbox[t]{2mm}{\multirow{8}{*}{\rotatebox[origin=c]{90}{PHY2}}}
 & MI-Iso
 & 0.2746 \newline p $<$ 1.00$\text{e-}$4
 & 0.2632 \newline p $<$ 1.00$\text{e-}$4
 & {\it \textcolor{gray!125} { 0.9889 \newline p = 0.858 }}
 & {\bf 0.4034 \newline p $<$ 1.00$\textbf{e-}$4 }
 & {\bf 0.0503 \newline p $<$ 1.00$\textbf{e-}$4 }
 \\ \cline{2-7}
 & IsoRankN
 & {\it \textcolor{gray!125} { 0.0000 \newline p = 1.000 }}
 & {\it \textcolor{gray!125} { 0.0000 \newline p = 1.000 }}
 & {\it \textcolor{gray!125} { 0.9863 \newline p = 0.382 }}
 & {\it \textcolor{gray!125} { 0.2521 \newline p = 0.009 }}
 & {\it \textcolor{gray!125} { 0.0067 \newline p = 0.382 }}
 \\ \cline{2-7}
 & GEDEVO-M
 & {\bf 0.5562 \newline p $<$ 1.00$\textbf{e-}$4 }
 & 0.7369 \newline p $<$ 1.00$\text{e-}$4
 & {\it \textcolor{gray!125} { 0.9786 \newline p = 0.042 }}
 & 0.2096 \newline p $<$ 1.00$\text{e-}$4
 & 0.0266 \newline p $<$ 1.00$\text{e-}$4
 \\ \cline{2-7}
 & multiMAGNA++
 & 0.5236 \newline p $<$ 1.00$\text{e-}$4
 & {\bf 0.7680 \newline p $<$ 1.00$\textbf{e-}$4 }
 & {\bf 0.9745 \newline p = 1.00\textbf{e-}4 }
 & {\it \textcolor{gray!125} { 0.1952 \newline p = 0.001 }}
 & 0.0249 \newline p = 1.00\text{e-}4
 \\ \hline
\parbox[t]{2mm}{\multirow{8}{*}{\rotatebox[origin=c]{90}{Y2H1}}}
 & MI-Iso
 & 0.1935 \newline p $<$ 1.00$\text{e-}$4
 & 0.0264 \newline p $<$ 1.00$\text{e-}$4
 & {\it \textcolor{gray!125} { 0.9229 \newline p = 0.995 }}
 & {\bf 0.2092 \newline p $<$ 1.00$\textbf{e-}$4 }
 & {\bf 0.0382 \newline p $<$ 1.00$\textbf{e-}$4 }
 \\ \cline{2-7}
 & IsoRankN
 & 0.0815 \newline p $<$ 1.00$\text{e-}$4
 & {\it \textcolor{gray!125} { 0.0000 \newline p = 1.000 }}
 & {\bf 0.8447 \newline p $<$ 1.00$\textbf{e-}$4 }
 & {\it \textcolor{gray!125} { 0.1736 \newline p = 0.304 }}
 & 0.0255 \newline p $<$ 1.00$\text{e-}$4
 \\ \cline{2-7}
 & GEDEVO-M
 & 0.4511 \newline p $<$ 1.00$\text{e-}$4
 & 0.0722 \newline p $<$ 1.00$\text{e-}$4
 & {\it \textcolor{gray!125} { 0.9032 \newline p = 0.919 }}
 & 0.1879 \newline p = 6.00\text{e-}4
 & 0.0347 \newline p $<$ 1.00$\text{e-}$4
 \\ \cline{2-7}
 & multiMAGNA++
 & {\bf 0.4943 \newline p $<$ 1.00$\textbf{e-}$4 }
 & {\bf 0.1088 \newline p $<$ 1.00$\textbf{e-}$4 }
 & 0.8899 \newline p = 6.00\text{e-}4
 & {\it \textcolor{gray!125} { 0.1794 \newline p = 0.124 }}
 & {\it \textcolor{gray!125} { 0.0322 \newline p = 0.719 }}
 \\ \hline
\parbox[t]{2mm}{\multirow{8}{*}{\rotatebox[origin=c]{90}{Y2H2}}}
 & MI-Iso
 & 0.2972 \newline p $<$ 1.00$\text{e-}$4
 & 0.0970 \newline p $<$ 1.00$\text{e-}$4
 & {\it \textcolor{gray!125} { 0.9880 \newline p = 0.313 }}
 & {\it \textcolor{gray!125} { 0.5567 \newline p = 0.131 }}
 & {\bf 0.0710 \newline p $<$ 1.00$\textbf{e-}$4 }
 \\ \cline{2-7}
 & IsoRankN
 & {\it \textcolor{gray!125} { 0.0000 \newline p = 1.000 }}
 & {\it \textcolor{gray!125} { 0.0000 \newline p = 1.000 }}
 & {\it \textcolor{gray!125} { 0.9885 \newline p = 0.641 }}
 & {\it \textcolor{gray!125} { 0.3333 \newline p = 0.870 }}
 & {\it \textcolor{gray!125} { 0.0022 \newline p = 0.214 }}
 \\ \cline{2-7}
 & GEDEVO-M
 & 0.6945 \newline p $<$ 1.00$\text{e-}$4
 & 0.1905 \newline p $<$ 1.00$\text{e-}$4
 & {\it \textcolor{gray!125} { 0.9968 \newline p = 0.926 }}
 & {\it \textcolor{gray!125} { 0.2228 \newline p = 0.714 }}
 & {\it \textcolor{gray!125} { 0.0303 \newline p = 0.486 }}
 \\ \cline{2-7}
 & multiMAGNA++
 & {\bf 0.7307 \newline p $<$ 1.00$\textbf{e-}$4 }
 & {\bf 0.2727 \newline p $<$ 1.00$\textbf{e-}$4 }
 & {\it \textcolor{gray!125} { 0.9926 \newline p = 0.271 }}
 & {\it \textcolor{gray!125} { 0.2342 \newline p = 0.354 }}
 & {\it \textcolor{gray!125} { 0.0328 \newline p = 0.069 }}
 \\ \hline
\end{tabular}
\normalsize
\end{center}
\caption{Alignment accuracy of different MNA approaches for the PHY1, PHY2, Y2H1 and Y2H2 network sets in terms of topological NCV-CIQ and LCCS measures and functional MNE, GO correctness (GC), and F-score measures, for topology-only alignments. The symbol ``p'' signifies $p$-values of the observed alignment scores. For each alignment quality measure, and for each network set, the best method (i.e., the method with the lowest $p$-value, or the method with the best alignment quality score if the $p$-values are tied) is bolded. The alignment scores that are not statistically significant, if any, are greyed out and italicized. Note that for MNE, the lower the score, the better the alignment quality. For all other measures, the higher the score, the better the alignment quality.}
\label{tab:all_all_top}
\end{table}

\begin{table}[h!]
\begin{center}
\scriptsize
\begin{tabular}{|l|p{22mm}|p{18mm}|p{18mm}|p{18mm}|p{18mm}|p{18mm}|} \hline
 & & \multicolumn{2}{c|}{Topological measures} & \multicolumn{3}{c|}{Functional measures} \\ \hline
 & Method
 & NCV-CIQ
 & LCCS
 & MNE
 & GC
 & F-score
 \\ \hline
\parbox[t]{2mm}{\multirow{10}{*}{\rotatebox[origin=c]{90}{PHY1}}}
 & MI-Iso
 & 0.2517 \newline p $<$ 1.00$\text{e-}$4
 & {\it \textcolor{gray!125} { 0.0000 \newline p = 1.000 }}
 & {\bf 0.8224 \newline p $<$ 1.00$\textbf{e-}$4 }
 & 0.2732 \newline p $<$ 1.00$\text{e-}$4
 & 0.0492 \newline p $<$ 1.00$\text{e-}$4
 \\ \cline{2-7}
 & IsoRankN
 & 0.1336 \newline p = 1.00\text{e-}4
 & 0.0258 \newline p $<$ 1.00$\text{e-}$4
 & {\it \textcolor{gray!125} { 0.8773 \newline p = 1.000 }}
 & {\bf 0.3279 \newline p $<$ 1.00$\textbf{e-}$4 }
 & {\bf 0.0838 \newline p $<$ 1.00$\textbf{e-}$4 }
 \\ \cline{2-7}
 & BEAMS
 & 0.3250 \newline p $<$ 1.00$\text{e-}$4
 & {\it \textcolor{gray!125} { 0.0000 \newline p = 1.000 }}
 & {\it \textcolor{gray!125} { 0.8944 \newline p = 0.895 }}
 & {\it \textcolor{gray!125} { 0.4084 \newline p = 1.000 }}
 & 0.0457 \newline p $<$ 1.00$\text{e-}$4
 \\ \cline{2-7}
 & FUSE
 & 0.0679 \newline p = 5.00\text{e-}4
 & {\it \textcolor{gray!125} { 0.0000 \newline p = 1.000 }}
 & 0.8781 \newline p $<$ 1.00$\text{e-}$4
 & 0.2268 \newline p $<$ 1.00$\text{e-}$4
 & 0.0472 \newline p $<$ 1.00$\text{e-}$4
 \\ \cline{2-7}
 & multiMAGNA++
 & {\bf 0.3884 \newline p $<$ 1.00$\textbf{e-}$4 }
 & {\bf 0.1064 \newline p $<$ 1.00$\textbf{e-}$4 }
 & 0.8622 \newline p $<$ 1.00$\text{e-}$4
 & 0.1759 \newline p $<$ 1.00$\text{e-}$4
 & 0.0353 \newline p $<$ 1.00$\text{e-}$4
 \\ \hline
\parbox[t]{2mm}{\multirow{10}{*}{\rotatebox[origin=c]{90}{PHY2}}}
 & MI-Iso
 & 0.1670 \newline p $<$ 1.00$\text{e-}$4
 & 0.1537 \newline p = 4.00\text{e-}4
 & 0.9736 \newline p = 3.00\text{e-}4
 & 0.2807 \newline p $<$ 1.00$\text{e-}$4
 & 0.0381 \newline p $<$ 1.00$\text{e-}$4
 \\ \cline{2-7}
 & IsoRankN
 & {\it \textcolor{gray!125} { 0.0000 \newline p = 1.000 }}
 & {\it \textcolor{gray!125} { 0.0000 \newline p = 1.000 }}
 & {\it \textcolor{gray!125} { 1.0000 \newline p = 0.376 }}
 & {\it \textcolor{gray!125} { 0.0000 \newline p = 1.000 }}
 & {\it \textcolor{gray!125} { 0.0000 \newline p = 1.000 }}
 \\ \cline{2-7}
 & BEAMS
 & {\bf 0.5858 \newline p $<$ 1.00$\textbf{e-}$4 }
 & 0.4714 \newline p $<$ 1.00$\text{e-}$4
 & {\bf 0.9799 \newline p $<$ 1.00$\textbf{e-}$4 }
 & {\it \textcolor{gray!125} { 0.5833 \newline p = 1.000 }}
 & 0.0210 \newline p $<$ 1.00$\text{e-}$4
 \\ \cline{2-7}
 & FUSE
 & {\it \textcolor{gray!125} { 0.0561 \newline p = 0.002 }}
 & {\it \textcolor{gray!125} { 0.0352 \newline p = 0.249 }}
 & {\it \textcolor{gray!125} { 0.9808 \newline p = 0.271 }}
 & 0.2422 \newline p $<$ 1.00$\text{e-}$4
 & 0.0332 \newline p $<$ 1.00$\text{e-}$4
 \\ \cline{2-7}
 & multiMAGNA++
 & 0.5226 \newline p $<$ 1.00$\text{e-}$4
 & {\bf 0.7578 \newline p $<$ 1.00$\textbf{e-}$4 }
 & 0.9748 \newline p = 2.00\text{e-}4
 & {\bf 0.3033 \newline p $<$ 1.00$\textbf{e-}$4 }
 & {\bf 0.0466 \newline p $<$ 1.00$\textbf{e-}$4 }
 \\ \hline
\parbox[t]{2mm}{\multirow{10}{*}{\rotatebox[origin=c]{90}{Y2H1}}}
 & MI-Iso
 & 0.1354 \newline p $<$ 1.00$\text{e-}$4
 & {\it \textcolor{gray!125} { 0.0000 \newline p = 1.000 }}
 & {\it \textcolor{gray!125} { 0.8992 \newline p = 0.396 }}
 & 0.1990 \newline p $<$ 1.00$\text{e-}$4
 & 0.0379 \newline p $<$ 1.00$\text{e-}$4
 \\ \cline{2-7}
 & IsoRankN
 & 0.1315 \newline p $<$ 1.00$\text{e-}$4
 & 0.0264 \newline p $<$ 1.00$\text{e-}$4
 & {\it \textcolor{gray!125} { 0.9003 \newline p = 0.031 }}
 & 0.3247 \newline p $<$ 1.00$\text{e-}$4
 & 0.0822 \newline p $<$ 1.00$\text{e-}$4
 \\ \cline{2-7}
 & BEAMS
 & 0.2856 \newline p $<$ 1.00$\text{e-}$4
 & {\it \textcolor{gray!125} { 0.0000 \newline p = 1.000 }}
 & {\it \textcolor{gray!125} { 0.9159 \newline p = 0.363 }}
 & {\bf 0.3945 \newline p $<$ 1.00$\textbf{e-}$4 }
 & {\bf 0.0856 \newline p $<$ 1.00$\textbf{e-}$4 }
 \\ \cline{2-7}
 & FUSE
 & 0.0480 \newline p $<$ 1.00$\text{e-}$4
 & {\it \textcolor{gray!125} { 0.0000 \newline p = 1.000 }}
 & {\bf 0.8781 \newline p $<$ 1.00$\textbf{e-}$4 }
 & 0.2369 \newline p $<$ 1.00$\text{e-}$4
 & 0.0483 \newline p $<$ 1.00$\text{e-}$4
 \\ \cline{2-7}
 & multiMAGNA++
 & {\bf 0.4383 \newline p $<$ 1.00$\textbf{e-}$4 }
 & {\bf 0.0722 \newline p $<$ 1.00$\textbf{e-}$4 }
 & {\it \textcolor{gray!125} { 0.8954 \newline p = 0.062 }}
 & 0.2040 \newline p $<$ 1.00$\text{e-}$4
 & 0.0428 \newline p $<$ 1.00$\text{e-}$4
 \\ \hline
\parbox[t]{2mm}{\multirow{10}{*}{\rotatebox[origin=c]{90}{Y2H2}}}
 & MI-Iso
 & 0.2035 \newline p $<$ 1.00$\text{e-}$4
 & 0.0773 \newline p = 2.00\text{e-}4
 & {\it \textcolor{gray!125} { 0.9913 \newline p = 0.604 }}
 & {\it \textcolor{gray!125} { 0.4554 \newline p = 0.783 }}
 & {\it \textcolor{gray!125} { 0.0693 \newline p = 0.082 }}
 \\ \cline{2-7}
 & IsoRankN
 & 0.2853 \newline p $<$ 1.00$\text{e-}$4
 & 0.0965 \newline p $<$ 1.00$\text{e-}$4
 & {\it \textcolor{gray!125} { 0.9945 \newline p = 0.704 }}
 & {\bf 0.3651 \newline p $<$ 1.00$\textbf{e-}$4 }
 & {\bf 0.0706 \newline p $<$ 1.00$\textbf{e-}$4 }
 \\ \cline{2-7}
 & BEAMS
 & 0.6020 \newline p $<$ 1.00$\text{e-}$4
 & {\bf 0.4741 \newline p $<$ 1.00$\textbf{e-}$4 }
 & {\it \textcolor{gray!125} { 0.9869 \newline p = 0.057 }}
 & {\it \textcolor{gray!125} { 0.8222 \newline p = 0.783 }}
 & 0.0616 \newline p $<$ 1.00$\text{e-}$4
 \\ \cline{2-7}
 & FUSE
 & {\it \textcolor{gray!125} { 0.0707 \newline p = 0.028 }}
 & {\it \textcolor{gray!125} { 0.0518 \newline p = 0.889 }}
 & {\it \textcolor{gray!125} { 0.9949 \newline p = 0.644 }}
 & 0.2801 \newline p $<$ 1.00$\text{e-}$4
 & 0.0416 \newline p $<$ 1.00$\text{e-}$4
 \\ \cline{2-7}
 & multiMAGNA++
 & {\bf 0.7125 \newline p $<$ 1.00$\textbf{e-}$4 }
 & 0.2361 \newline p $<$ 1.00$\text{e-}$4
 & {\it \textcolor{gray!125} { 0.9908 \newline p = 0.090 }}
 & 0.2920 \newline p $<$ 1.00$\text{e-}$4
 & 0.0436 \newline p $<$ 1.00$\text{e-}$4
 \\ \hline
\end{tabular}
\normalsize
\end{center}
\caption{Alignment accuracy of different MNA approaches for the PHY1, PHY2, Y2H1 and Y2H2 network sets in terms of topological NCV-CIQ and LCCS measures and functional MNE, GO correctness (GC), and F-score measures, for topology+sequence alignments. The symbol ``p'' signifies $p$-values of the observed alignment scores. For each alignment quality measure, and for each network set, the best method (i.e., the method with the lowest $p$-value, or the method with the best alignment quality score if the $p$-values are tied) is bolded. The alignment scores that are not statistically significant, if any, are greyed out and italicized. Note that for MNE, the lower the score, the better the alignment quality. For all other measures, the higher the score, the better the alignment quality.}
\label{tab:all_all_topseq}
\end{table}

\clearpage

%
%

\bibliographystyle{natbib}
\bibliography{cone}

\end{document}